\documentclass[twocolumn]{aastex7}
\usepackage{hyperref}
\usepackage{bm}
\usepackage{amsmath}
\usepackage{booktabs}

\newcommand{\be}{\begin{equation}}
\newcommand{\ee}{\end{equation}}
\newcommand{\msun}{M_\odot}

\begin{document}

\shorttitle{Ultra Heavy Cosmic Rays from Magnetars}
\shortauthors{Patel, Diesing, \& Metzger}

\title{Ultra Heavy Cosmic Rays from Magnetars}

\author[0009-0000-1335-4412]{Anirudh Patel}
\affil{Department of Physics and Columbia Astrophysics Laboratory, Columbia University, New York, NY 10027, USA}
\email[show]{ap4268@columbia.edu}  

\author[0000-0002-6679-0012]{Rebecca Diesing}
\affil{Department of Physics and Columbia Astrophysics Laboratory, Columbia University, New York, NY 10027, USA}
\affil{School of Natural Sciences, Institute for Advanced Study, Princeton, NJ 08540, USA}
\email[]{rrd2148@columbia.edu}  

\author[0000-0002-4670-7509]{Brian D.~Metzger}
\affil{Department of Physics and Columbia Astrophysics Laboratory, Columbia University, New York, NY 10027, USA}
\affil{Center for Computational Astrophysics, Flatiron Institute, 162 5th Ave, New York, NY 10010, USA} 
\email[]{bdm2129@columbia.edu}

\begin{abstract}
Matter ejected from the neutron star crust during a magnetar giant flare will undergo $r$-process nucleosynthesis during decompression. Ultra heavy ions ($Z \gg 26$) can be accelerated to cosmic ray energies by the reverse shock as the ejecta decelerates by interacting with the ambient environment. We investigate the contribution of magnetars to the local ultra heavy cosmic ray flux using semi-analytic Galactic transport calculations, demonstrating that they may be significant contributors throughout Galactic history depending on the giant flare rate and ion acceleration efficiency. Although neutron star mergers inject orders of magnitude more energy into cosmic rays, they rarely occur within the spallation-limited propagation horizon for ultra heavy species, reducing their local contributions. As compared to lighter nuclei which are dominantly accelerated by supernovae, the SuperTIGER experiment has presented tentative evidence for a distinct contribution to the cosmic ray abundances near and above the first $r$-process peak ($Z \approx 35\text{--}56$).  We argue that current abundance data are consistent with either a magnetar giant flare or neutron star merger origin for these species. Measurements with single element resolution through the third $r$-process peak, expected from the upcoming TIGERISS experiment, may discriminate between these sources for the heaviest cosmic rays. 

\end{abstract}

\section{Introduction} 

Magnetars exhibit a variety of transient and persistent emission powered by their extreme magnetic fields \citep{Thompson&Duncan1995, Kaspi&Beloborodov2017}. Their remarkable giant flares, sub-second gamma-ray bursts with extended ($\sim 100~\rm s$) hard X-ray tails, rank amongst the brightest extrasolar transients ever recorded \citep{Mazets+1979,Hurley+05,Palmer+2005}. Synchrotron radio afterglows observed in the days following the flares reveal that these events are accompanied by baryonic ejecta from the neutron star \citep{Gelfand+2005, Granot+2006}. 

Baryon loaded outflows from magnetar giant flares (MGFs) have recently gained renewed interest as novel sites of rapid neutron capture process ($r$-process) nucleosynthesis (\citealt{Cehula+24,patel+25a, Patel+2025b}). These works demonstrated that the delayed MeV signal observed after the 2004 MGF from SGR 1806-20 \citep{Mereghetti+05,Boggs+2007,Frederiks+07} is naturally explained by nuclear line emission from newly synthesized $r$-process nuclei. This distinct ejecta composition implicates MGFs as potential sources of $r$-process element cosmic rays \citep{patel+25a}.

The chemical and isotopic composition of Galactic cosmic rays offers a unique probe of particle accelerators, their ambient environments, and the microphysics of acceleration and transport. For instance, the systematic overabundance of refractory elements relative to volatiles has motivated models in which ionized dust grains—and the embedded refractory elements—are preferentially accelerated at shocks due to their high rigidity \citep{Epstein1980, Meyer+1997, Ellison+1997, Cristofari+2025}. These fractionation patterns, combined with the unusually large $^{22}\rm Ne/^{20}Ne$ ratio and other isotopic anomalies, indicate that a significant fraction of cosmic rays originate in superbubble environments of nominal solar-like ISM composition enriched $\sim 20\%$ by mass with Wolf-Rayet wind/ejecta (e.g., \citealt{Garcia-Munoz+1979, Higdon&Lingenfelter2003, Binns+2005, Binns+2016}). An $80\%$ ISM and $20\%$ massive-star material source composition indeed yields a clean separation between refractory and volatile elements in abundance patterns, supporting this ``OB association" model for the Galactic cosmic ray source \citep{Rauch+2009, Murphy+2016}.\footnote{The OB association is further supported by hadronic emission observed around young stellar clusters (e.g., \citealt{Ackermann+2011, Aharonian+2019}).}

While a volatility-based acceleration model coupled with an OB association source successfully explains the abundances of most Galactic cosmic rays, measurements from the SuperTIGER experiment suggest this paradigm may be insufficient for the nuclear species much heavier than iron, {\it ultra heavy cosmic rays} (UHCRs, \citealt{Walsh+2023}). The tension arises near the first $r$-process peak ($Z \sim 35{\text{--}}40$) and extends through the second peak at $Z=56$ (the highest charge currently reported; see Fig.~\ref{fig:abund}). 
The breakdown of the volatility-based acceleration model in this regime, coincident with apparent abundance enhancements, implicates a distinct UHCR source population: one that is highly enriched in $r$-process species and depleted in dust relative to the standard ISM/OB source material.

These conditions may naturally be satisfied in $r$-process dominated outflows from nucleosynthesis events. In neutron star merger (NSM) ejecta, the high expansion velocities $v_{\rm ej} \gtrsim 0.1c$ and relative depletion of alpha-group elements (which dominate nucleation seeds) render dust formation inefficient; indeed, the near-infrared excess in kilonova photometry is inconsistent with emission arising from substantial dust reprocessing \citep{Gall+17}. MGF ejecta share kinematic and compositional properties with NSMs, but are even less dense because of their lower masses, implying they are unlikely to form dust prior to interacting with the external medium. MGFs and NSMs are therefore both plausible candidate sources for the UHCR anomaly reported by SuperTIGER. \citet{Komiya&Shigeyama2017} demonstrated that NSMs may contribute to the UHCR flux over Galactic history, however, the contribution from MGFs has not been investigated.

In this Letter we solve the time-dependent diffusion-loss equation to predict the local ultra heavy cosmic ray flux contributions from magnetar giant flares and neutron star mergers. We characterize these source populations in Section~\ref{sec:sources} and summarize our model, which builds upon that of \citet{Komiya&Shigeyama2017}, in Section~\ref{sec:model}. In Section~\ref{sec:local_flux} we present our simulation results comparing the time-dependent flux from different sources. We examine these in the context of available UHCR data and offer interpretations for the anomalous measurements in Section~\ref{sec: abund}. We summarize our results and briefly discuss broader implications in Section~\ref{sec:conclude}.

\section{Sources of Ultra Heavy Cosmic Rays}
\label{sec:sources}
\begin{table*}
\centering
\caption{Sources of ultra heavy cosmic rays.}
\begin{tabular}{ccccccccc}
\toprule
Accelerator (Shock) & Accelerated Medium & $X_r$ & $E_{\rm ej}~\rm(erg)$ & $M_{\rm ej}~\rm(\msun)$ & $v_{\rm ej}~\rm(c)$ & $\mathcal{R}~\rm(Myr^{-1})$ & Dust? \\
\midrule
SN (FS) & ISM/OB & $\sim 10^{-7}$ & $10^{51}$ & $\geq1$ & $0.01\text{--}0.05$ & $1\text{--}3 \times 10^4$ & Y \\
MGF (RS) & MGF ejecta & $\sim 1$ & $10^{44\text{--}47}$ &  $10^{-8\text{--}6}$ & $0.1\text{--}0.7$ & $\sim 10^{3\text{--}4}$ & N \\
NSM (RS) & NSM ejecta & $\sim 1$ & $10^{51}$ &  $10^{-2}$ & $0.1\text{--}0.3$ & $\sim 1\text{--}20$ & N \\
\bottomrule
\end{tabular}
\label{tab:sources}
\end{table*}

\begin{figure*} 
    \centering
    \includegraphics[width=1.0\textwidth]{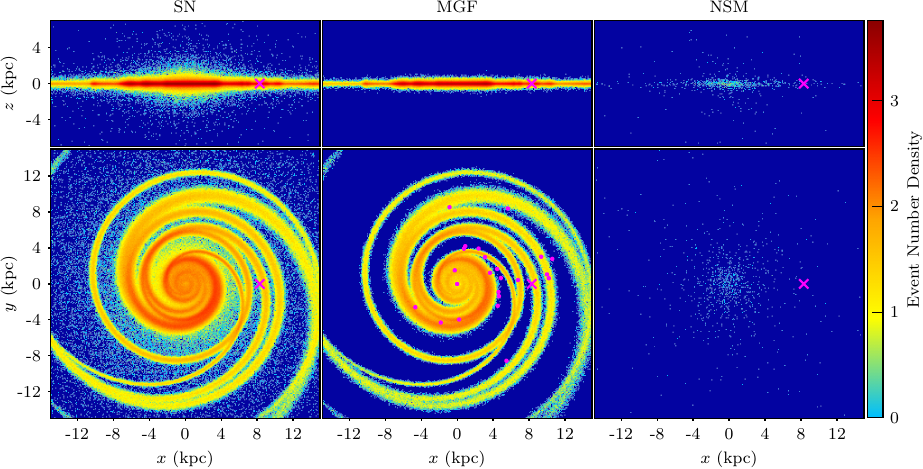}
    \caption{Surface density map of Milky Way SNe, MGFs, and NSMs stochastically generated over $50~\rm Myr$ with fiducial rates $\mathcal{R}_{\rm SN} = 3\times10^4~\rm Myr^{-1}$, $\mathcal{R}_{\rm MGF} = 1.2\times10^4~\rm Myr^{-1}$ and $\mathcal{R}_{\rm NSM} = 20~\rm Myr^{-1}$, where the MGF rate includes only the most energetic events, $E_{\rm iso} \geq 10^{46}~\rm erg$. The locations of 24 Galactic magnetars are marked as pink dots in the middle panel \citep{Olausen&Kaspi2014}. The Solar System is located at $\bm{r}_\odot =8.3\bm{\hat{x}}$ denoted by the pink ``x". The color bar is logarithmically scaled.} 
    \label{fig:sources}
    
\end{figure*}

Heavy ions may be injected into the Galactic UHCR population promptly (on a Sedov timescale) following ejection from their nucleosynthetic sites. This generically occurs through diffusive shock acceleration (DSA, \citealt{Krymskii1977, Axford+1977, Bell1978a, Blandford&Ostriker1978}) at the reverse shock propagating through the $r$-process enriched ejecta as it decelerates upon interaction with the large-scale gaseous environment surrounding the magnetar. The rate at which energy is injected into the UHCR population is $\dot{E}_{\rm CR} \propto X_rE_{\rm ej}\mathcal{R}$, where $X_r$ is the mass fraction of $r$-process nuclei in the accelerated medium, $E_{\rm ej}$ is the ejecta energy (a small fraction of which is transferred to cosmic rays, detailed in Sec.~\ref{sec:model}), and $\mathcal{R}$ is the event rate. By this metric alone, NSMs, with $\mathcal{R}_{\rm NSM} = 1 \text{--} 20~\rm Myr^{-1}$ \citep{ligo_cat_4_2025}, $E_{\rm ej} \sim 10^{51}~\rm erg$ (e.g., \citealt{cowperthwaite2017electromagnetic}), and $X_r\sim 1$, should be the dominant sources of UHCRs in the Galaxy. However, the local interstellar flux from NSMs is severely attenuated by propagation losses, particularly spallation, due to the low merger rate and corresponding long propagation timescales for particles to reach the Solar neighborhood \citep[see also Sec.~\ref{sec:model}]{Komiya&Shigeyama2017}. 

More frequent events like MGFs may then contribute significantly to the local flux relative to NSMs. Three Galactic (or Magellanic) MGFs have been observed in the last $50$ years with a range of isotropic equivalent energy in their prompt emission $E_{\rm iso} \sim 10^{44 \text{--}46}~\rm erg$  \citep{Mazets+1979, Hurley+1999, Palmer+2005, Hurley+05}. The kinetic energy of the baryonic ejecta is expected to be similar to the energy of the flare itself. For the 2004 flare from SGR 1806-20, constraints on the ejecta mass and velocity inferred from nonthermal radioactive emission require a kinetic energy $E_{\rm ej} \geq 10^{46}~\rm erg$ \citep{patel+25a}, consistent with that inferred from radio synchrotron emission \citep{Gelfand+2005, Granot+2006}. The Galactic rate of these powerful SGR 1806-20-like flares is currently estimated at $\mathcal{R} \approx 1.3^{+1.2}_{-1.0} \times 10^4~\rm Myr^{-1}$ (90\% confidence, \citealt{Burns+2021}). This is determined from the total rate and inferred intrinsic energy distribution $\propto E_{\rm iso}^{-1.3}$; weaker flares are more common but release less total energy in aggregate. Like NSMs, the baryon ejecta of MGFs is highly enriched in $r$-process nuclei $X_r\sim 1$ \citep{Patel+2025b}.

UHCRs need not be injected exclusively by $r$-process events. The ambient ISM and OB association source comprise a small (solar-like) mass fraction of $r$-process nuclei $X_r\sim 10^{-7}$ accumulated over many nucleosynthetic events. These nuclei can be injected into the UHCR population at forward shocks. Supernovae (SNe) will overwhelmingly dominate this injection channel as their large explosion energies $E_{\rm ej} \sim 10^{51} ~\rm erg$ and high rates $\mathcal{R} = 3 \times10^4~\rm Myr^{-1}$ compensate for the trace levels of $r$-process species in their ambient environments. We thus consider SNe, MGFs, and NSMs as promising sources of UHCRs in the Milky Way. Their properties are summarized in Table~\ref{tab:sources}.

\section{Model}
\label{sec:model}
We model the local UHCR flux resulting from source injection by SNe, MGFs, and NSMs accounting for Galactic transport (diffusion, energy loss, fragmentation) and local modulation effects as described in \citet{Komiya&Shigeyama2017}. Critically, our model accounts for particle propagation in the Galactic disk and halo with free escape at the boundaries, and uses standard transport parameters constrained by observations. We also incorporate a realistic spatial distribution of sources generated through a Monte Carlo procedure (Appendix~\ref{sec:rsg}) with fiducial event rates $\mathcal{R}_{\rm SN} = 3 \times 10^4~\rm Myr^{-1}$, $\mathcal{R}_{\rm MGF} = 1.2 \times 10^4 ~\rm Myr^{-1}$, and $\mathcal{R}_{\rm NSM} = 20~\rm Myr^{-1}$. A density map of events simulated over $50~\rm Myr$ with these rates is shown in Fig.~\ref{fig:sources}. These sources inject different nuclear species into the UHCR population with abundances (unique to the source type) derived from observations or nuclear reaction network calculations. Main features of the model are summarized below, with details presented in the Appendix.

\subsection{Injection}

The source spectrum of a species of mass $m$ injected through DSA can be approximated as a power law with a high energy exponential cutoff,
\be
N_0(E) = \mathcal{N} \bigg( \frac{E}{0.1mc^2}\bigg)^{-q} e^{-E/E_{\rm max}},
\label{eq:spectrum}
\ee
where we assume the description applies for particles accelerated to Lorentz factors $\gamma \geq 1.1$ (kinetic energy $ \varepsilon \equiv E/A \gtrsim 100~\rm MeV~A^{-1}$). Nonlinear DSA accounting for downstream drift of magnetic scattering centers predicts steep spectra $q = 2.4$ \citep[Appendix~\ref{sec:injection}]{Diesing&Caprioli2021}. The spectral cut off is set by the maximum energy $E_{\rm max}$ achieved by particles in DSA, which is determined by the size of the acceleration region \citep{Diesing2023},
\begin{multline}
E_{\rm max} \approx Z \times 10^{15}\,{\rm eV} \\
\times  \left(\frac{\xi_{\rm CR}}{0.1}\right)^{1/2}
\left(\frac{n_0}{1~{\rm cm^{-3}}}\right)^{1/6}
\left(\frac{v_{\rm ej}}{0.2c}\right)^{11/6}
\left(\frac{E_{\rm ej}}{10^{46}~\rm erg}\right)^{1/3},
\label{eq:Emax}
\end{multline}
where $\xi_{\rm CR}$ is the cosmic ray injection efficiency, $n_0$ is the upstream density, and $v_{\rm ej}$ and $E_{\rm ej}$ are the ejecta velocity and kinetic energy (Table~\ref{tab:sources}).

The normalization $\mathcal{N}$ is proportional to the injection efficiency $\xi_{\rm CR} \equiv E_{\rm CR}/E_{\rm ej}$ where $E_{\rm CR}$ is the energy transferred to the cosmic ray species from the ejecta. Kinetic simulations reveal alpha-group ions are preferentially entrained in DSA, resulting in an enhanced injection efficiency proportional to the mass to charge ratio $(A/Z)^2$ \citep{Caprioli+2017, Caprioli+2025}. We assume this scaling extends to ultra heavy ion injection and adopt $\xi_{\rm CR} = \eta X(A/Z)^2$, where $X$ is the mass fraction of the species in the accelerated medium. We use a forward shock efficiency normalization $\eta_{\rm FS} = 0.15$ which is constrained by the Galactic cosmic ray proton energy budget \citep{Reynolds2008} and supported by observations and kinetic simulations \citep{Morlino&Caprioli2012, Caprioli&Spitkovsky2014}.

The acceleration efficiency of ions at reverse shocks is not well constrained for SNe, nor has it been studied for NSM or MGF outflows. 
While the maximum particle energy achieved through DSA depends sensitively on the amplified magnetic field strength around the shock front, the effective operation of DSA (to arbitrary energies) requires the presence of a {\it non-negligible} magnetic field.
Observations of synchrotron emitting regions and corresponding polarization measurements following the SGR 1806-20 MGF provide evidence for magnetic field structure in the ejecta \citep{Taylor+2005}, indicating that magnetic field amplification may be efficient. In such weakly magnetized shocks, magnetic fields can be amplified through the Weibel instability resulting in ion acceleration with efficiency $\sim 0.1$ \citep{Jikei+2025}. Given that a fraction $\lesssim 0.5$ of the ejecta energy is transferred to the reverse shock,  we use $\eta_{\rm RS} = 0.01$. We assume the same for reverse shocks in NSMs.

\subsection{Transport}

The differential number density $N$ of particles with kinetic energy $E$ at position $\bm{r}$ and time $t$ in the Galaxy is determined by solving the transport equation,
\be
\frac{\partial N(E, \bm{r}, t)}{\partial t} = \nabla \cdot (D\nabla N) - \frac{\partial (N\dot{E})}{\partial E} - N\Gamma_{\rm sp} + Q(E, \bm{r}, t).
\label{eq:transport}
\ee
We adopt a phenomenological spatial diffusion coefficient for isotropic scattering in interstellar magnetic turbulence $D = D_0\beta (R/1~\rm GV)^{\delta}$ for a particle of rigidity $R$ and velocity $\beta$ (normalized to the speed of light $c$). The normalization $D_0$ and rigidity scaling $\delta$ are constrained by observations (described in the next sub-section). Relativistic nuclei cool and fragment through electromagnetic and hadronic interactions in the interstellar medium. The particle cooling rate $\dot{E} (Z,E)$ includes contributions from ionization, Coulomb scattering, and pion production. The dominant losses are incurred through nuclear spallation at a rate $\Gamma_{\rm sp}(A, E)$ (Appendix~\ref{sec:decay}). 

The source term $Q = N_0(E)\delta(E-E_0) \delta^3(\bm{r}- \bm{r}_0)\delta(t-t_0)$ describes the impulsive injection of a differential particle spectrum $N_0(E)$ (Eq.~\ref{eq:spectrum}) from an event at coordinates $(t_0,\bm{r}_0)$. The cosmic ray density measured at Earth ($\bm{r}_\odot =8.3\bm{\hat{x}}~\rm kpc$) is then,
\be
N_\oplus(E, \bm{r}_\odot, t) = f_\odot(E)\int_0^\infty N_0(E_0) G(E+\Phi, \bm{r}_\odot, t) dE_0,
\label{eq:N_obs}
\ee
where $G$ is the Green's function for the solution to Eq.~\eqref{eq:transport} and the heliospheric suppression  $f_\odot(E)$ and energy shift $\Phi$ account for solar cycle modulation of the local interstellar spectrum as described in Appendix \ref{sec:decay}.

\begin{figure} 
    \centering
    \includegraphics[width=1.\columnwidth]{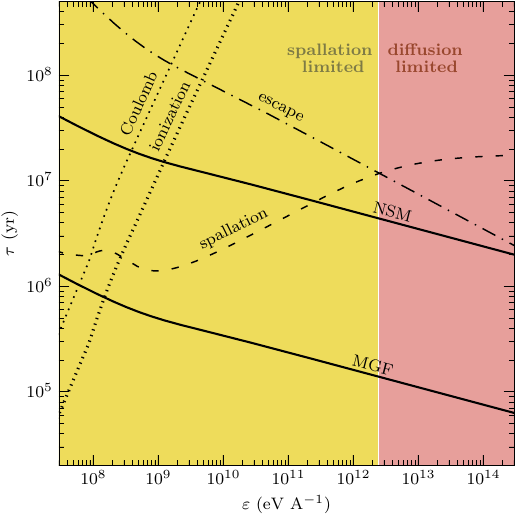}
    \caption{The energy loss, nuclear spallation, and escape timescales (patterned lines) for the local interstellar population of a representative ultra heavy species $^{136}$Xe, compared to the characteristic propagation ages of cosmic rays from NSMs and MGFs (solid lines). The shaded regions represent the energy regimes in which the propagation volume is limited by the spallation horizon (yellow) and the diffusion horizon (i.e., halo height, pink).}
    \label{fig:timescales}
\end{figure}

The characteristic propagation age of particles from a source within $d = 20~\rm kpc$ of the Solar neighborhood is $\tau_{\rm CR} \sim d(D\mathcal{R})^{-1/2}$ \citep{Komiya&Shigeyama2017}. The spallation timescale for GeV particles $1/\Gamma_{\rm sp} \sim 1~\rm Myr$ is much shorter than the propagation age of particles from NSMs (Fig.~\ref{fig:timescales}) such that spallation significantly suppresses the incident particle flux. The comparatively high MGF rate translates to short propagation ages, ensuring that UHCRs diffusing in the local environment are frequently ``replenished", rendering spallation less significant in setting the final flux. More intuitively, one can consider an effective spallation horizon $H_{\rm sp} =\sqrt{D\tau_{\rm sp}}$, the characteristic distance beyond which the incident particle flux is significantly attenuated by fragmentation. For ultra heavy species, $H_{\rm sp} < 1~\rm kpc$ such that NSMs rarely occur within the horizon. This enables MGFs (and SNe) to compete with NSMs as UHCR sources despite their lower particle injection rates.

\subsection{Calibrating Transport Parameters}
The mean gas density experienced by particles over their propagation volume is $\bar{n} \approx n_{\rm disk} h/H_{\rm eff}$ since particles diffuse in the disk ($|z| < h$) with $n_{\rm disk} = 1~\rm cm^{-3}$ and in the halo ($h < |z| < H_{\rm eff}$) with $n_{\rm halo} \ll n_{\rm disk}$ where propagation losses are comparatively negligible. The sample of particles measured at Earth ($z = 0$) have diffused over an effective height set by either their spallation horizon or the halo height, $H_{\rm eff} = H_{\rm sp}\tanh(H/H_{\rm sp})$. The spallation horizon is comparable to the disk height for ultra heavy species while the propagation volume for lighter species is not limited by spallation and spans the entire halo height. Particles are not diffusive for $z>H$ where they stream away or are advected by the Galactic wind, which is enforced by free escape boundary conditions $N(z = \pm H) = 0$ (Appendix~\ref{sec:decay}). Stated another way, the escape time $\tau_{\rm esc} = H^2/2D$ is less than the spallation timescale for light species, whereas one can identify the transition from spallation-limited transport to diffusion-limited transport ($\tau_{\rm esc} = \tau_{\rm sp}$) at energy $\varepsilon \sim 10^{12\text{--}13}~\rm eV$ for UHCRs (Fig.~\ref{fig:timescales}). We adopt a halo size $H = 7~\rm kpc$ motivated by radioisotope measurements \citep{Evoli+2020} and $h = 0.3~\rm kpc$. 

With the model geometry and effective ISM gas density set, we can utilize observational constraints to select the parameters $D_0$ and $\delta$ entering the diffusion coefficient. The grammage accumulated by a light cosmic ray species (which diffuses over the entire halo height) is $\mathcal{X} \approx m_p\bar{n}\beta cH^2D^{-1}$. The grammage is proportional to the secondary to primary particle ratio; fitting boron to carbon measurements yields $\mathcal{X}(R) \approx 12~{\rm g~cm^{-2}}\, (R/10~{\rm GV})^{-(2.85-q)}$ \citep{Blasi2017}. For an injection spectrum $q = 2.4$ (Eq.~\ref{eq:spectrum}), the normalization and scaling index for the diffusion coefficient are $D_0 = 2\times10^{28}~\rm cm^2~s^{-1}$ and $\delta = 2.85-q = 0.45$, respectively. 

Since we do not self-consistently track the secondary production chain, this calibration is approximate and is not intended to capture subtle spectral features across the full range of energies or nuclear species. However, as described in the next section, our model successfully reproduces major properties of observed spectra. We deem this sufficient for the present study, which aims to estimate relative contributions to the UHCR flux rather than resolve fine structure in the spectra.

\section{The Local Cosmic Ray Flux}
\label{sec:local_flux}

\begin{figure} 
    \centering
    \includegraphics[width=1.\columnwidth]{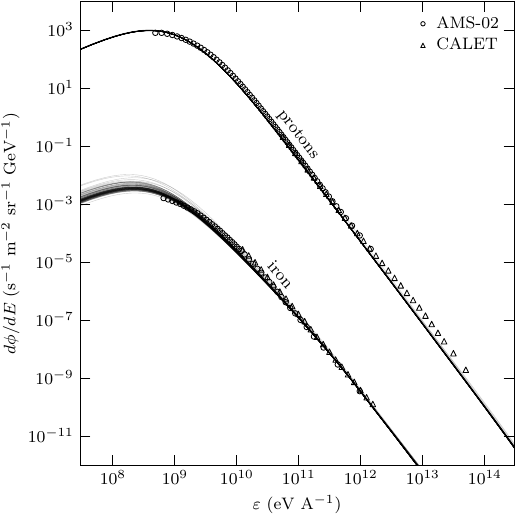}
\caption{Simulated proton and iron spectra for the fiducial SN rate $\mathcal{R} = 3\times10^4~\rm Myr^{-1}$ and kinetic energy $E_{\rm ej} =10^{51}~\rm erg$, compared to measurements from AMS-02 \citep{Aguilar+2015, Aguilar+2021} and CALET \citep{Adriani+2021, Adriani+2022}. Each thin line is a realization of the spectrum resulting from a stochastically generated source distribution (100 iterations total).}
\label{fig:spectrum}
\end{figure}

We first verify our model by comparing simulation results to the well measured cosmic ray proton and iron spectra, which are expected to be dominantly produced (below the knee) by SNe. For a fiducial rate assumed to include both core-collapse and thermonuclear SNe $\mathcal{R_{\rm SN}} = 3\times 10^4~\rm Myr^{-1}$ and typical ejecta energy $E_{\rm ej} = 10^{51}~\rm erg$, our calculations successfully reproduce the proton spectrum measured by AMS-02 and CALET (Fig.~\ref{fig:spectrum}). Our calculated iron spectrum requires a factor of $4$ enhancement in the acceleration efficiency to match the spectra, which is the expected refractory element enhancement reported by \citet{Murphy+2016}. We appropriately apply a factor of 2 enhancement to the UHCR flux from SNe (since about 50\% of the ISM/OB $r$-process composition is refractory), but not to MGFs or NSMs since their ejecta is depleted in dust.

\begin{figure} 
    \centering
    \includegraphics[width=1.0\columnwidth]{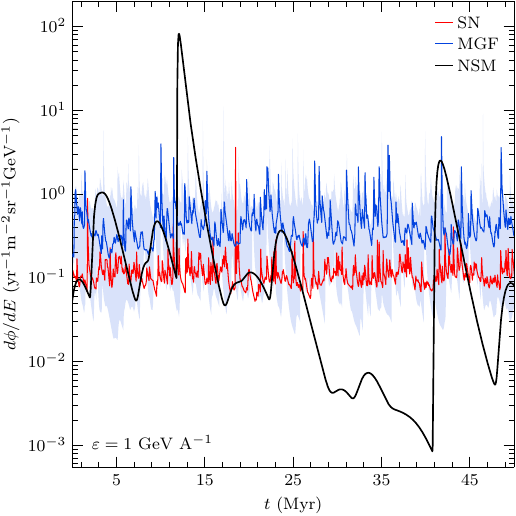}
    \caption{The local UHCR flux at $\varepsilon = 1~\rm GeV~A^{-1}$ from SNe, MGFs, and NSMs with fiducial parameters described in the text (solid lines). The light blue shaded region encompasses the UHCR flux for the upper and lower limit in the inferred MGF rate.}
\label{fig:flux}
\end{figure}

We simulate the UHCR flux of species with $Z\geq 40$, for which the injected mass fractions are $X_{r,\rm SN} = 10^{-7}$ , $X_{r,\rm MGF} = 0.6$, and $X_{r,\rm NSM} = 0.2$, respectively. We have assumed OB association source abundances for SNe \citep{Lodders+2020, Woosley+Heger2007}, nuclear network yields for MGF models shown to reproduce observed properties of the SGR 1806-20 flare \citep{patel+25a, Patel+2025b}, and a solar $r$-process composition for NSMs \citep{Prantzos+2020}. 

Fig.~\ref{fig:flux} presents the contributions to the local UHCR flux at $\varepsilon = 1~\rm GeV~A^{-1}$ for a representative Monte Carlo iteration with the fiducial MGF and NSM rates and energies. We find the persistent UHCR flux from MGFs to exceed that from SNe at most epochs, except in the event of a particularly nearby SN (e.g., $t=19~\rm Myr$). The flux from NSMs exhibits significant time-variability due to the low probability of nearby mergers and severe attenuation of the distant background by spallation (Fig.~\ref{fig:timescales}). Across 100 Monte Carlo realizations, we find NSMs dominate the local UHCR flux for a fraction $20^{+7}_{-9}\%$ of time ($90\%$ confidence).

We consider variations in source parameters around the fiducial values for MGFs and NSMs. The blue shaded region in Fig.~\ref{fig:flux} is bounded by the UHCR flux at the lower and upper limits in the inferred MGF rate, $\mathcal{R}_{\rm MGF} = 3 \times 10^3~\rm Myr^{-1}$ and $\mathcal{R}_{\rm MGF} = 2.5\times 10^4~\rm Myr^{-1}$. The median flux changes linearly with the rate, with the lower (higher) rate model exhibiting larger (smaller) amplitude fluctuations about the median. In the low rate limit, the bulk UHCR flux still competes with that from SNe at many epochs in time. The effect of decreasing the NSM rate from the fiducial value is to change the frequency and amplitude of the peaks in the flux (e.g., $12~\rm Myr$ in the fiducial model). For the lower limit reported by LIGO, $\mathcal{R}_{\rm NSM} = 1~\rm Myr^{-1}$, we find NSMs dominate the flux only $1^{+3}_{-1}\%$ of the time; for a rate $\mathcal{R}_{\rm NSM} = 10~\rm Myr^{-1}$, they dominate $10^{+5}_{-7}\%$ of the time. NSMs will become increasingly dominant at greater particle energies as the spallation horizon increases (Fig~\ref{fig:timescales}).

\section{Interpretation Of Observed Abundance Pattern}
\label{sec: abund}

\begin{figure} 
    \centering
    \includegraphics[width=1.0\columnwidth]{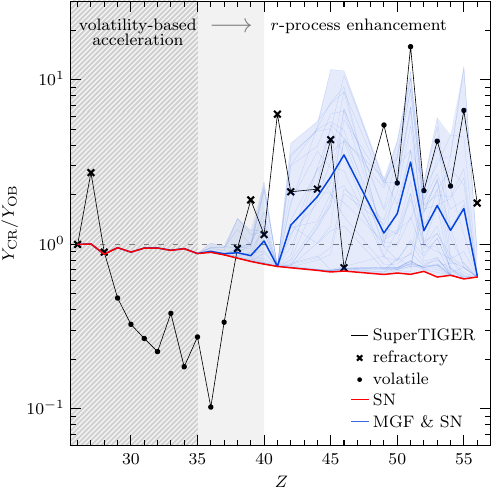}
\caption{ The preliminary UHCR abundance pattern measured by SuperTIGER (black) at $\varepsilon = 0.3 \text{--}10~\rm GeV~A^{-1}$ follows the standard OB association source model (80\% solar and 20\% massive-star material, \citealt{Murphy+2016}) up until $Z \sim 35 \text{--} 40$, above which an excess in the $r$-process abundances is coincident with a break-down of the volatility-based acceleration pattern (which we do not attempt to model).  This is consistent with a contribution from a non-SN cosmic ray source such as MGFs. Propagated cosmic ray abundances within our model from SNe only (red), and with the addition of MGFs (blue), are shown for comparison. Abundances are arbitrarily scaled to $\rm Y_{\rm CR}(Fe)/Y_{\rm OB}(Fe)=1$.}
\label{fig:abund}
\end{figure}

Our simulation results indicate that UHCRs from MGFs may produce relative abundance enhancements in broad agreement with the excess measured by SuperTIGER for $Z\gtrsim40$ \citep{Murphy+2016, Walsh+2023}. Fig.~\ref{fig:abund} shows the normalized abundance pattern in the local UHCR flux across the SuperTIGER energy band $ \varepsilon = 0.3 \text{--}10~\rm GeV~A^{-1}$ (averaged over randomly sampled times). The dark blue line includes contributions from both SNe and MGFs, and displays the propagated abundances averaged across $30$ MGF nucleosynthesis trajectories from \citet{Patel+2025b}. The blue shaded region spans the full range of abundances resulting from individual trajectories (thin blue lines). 

Were the UHCR anomaly to originate from a NSM, the incident particle flux would have to be comparable to that from SNe to produce the appropriate relative abundance enhancement demonstrated for the case of MGFs by Figs.~\ref{fig:flux} \& \ref{fig:abund}. This would preclude the possibility that we are receiving peak flux from a nearby ($\lesssim \rm few~kpc$) and recent ($\lesssim 5~\rm Myr$) NSM which would yield an orders-of-magnitude abundance enhancement (e.g., at $t = 12~\rm Myr$) not seen in the data. However, this does not exclude that the anomalous abundance pattern measured by SuperTIGER includes contributions from a NSM at a lower flux epoch (e.g., $t = 5~\rm Myr$).  A NSM component may even be required to explain the anomaly if the MGF contribution lies near the low end of its uncertainty range and hence fails to produce the full $r$-process enhancement.

Our analysis thus indicates that both MGFs and NSMs are viable source candidates for the UHCR anomaly observed by SuperTIGER. Current models of MGF nucleosynthesis show a steep drop (to mass fractions $X \lesssim 10^{-6}$) in the yields above $Z=60$ \citep{Patel+2025b} whereas NSM are expected to robustly produce elements through the third $r$-process peak at Pt ($Z = 78$; possibly accounting for much of the total synthesis of $r$-process elements, at least at the present epoch in Galactic history; e.g., \citealt{Hotokezaka+2018}).  Measurements with single element resolution have not been conducted for $Z > 56$ but some constraints in this regime are provided by HEAO-3 HNE and UHCRE measurements of element ``groups" \citep{Binns+1989, Donnelly+2012}. These suggest modest (factors of $\lesssim 4$) enhancements in the actinide/sub-actinide and Pt-group/Pb-group ratios relative to solar. However, the actinide and Pt $r$-process element groups are also highly refractory, making it challenging to discern whether the observed enhancements reflect acceleration at an $r$-process enriched source (i.e., NSM) or volatility-based acceleration at an ISM/OB source. Future measurements with single element resolution through the third $r$-process peak (e.g., \citealt{Zober+2025}) should help break this degeneracy and constrain whether a third peak enhancement is present in the UHCR abundances. 

We note the significant discrepancies in the detailed abundance patterns predicted by our MGF model suite and the SuperTIGER data. Such discrepancies do not themselves disfavor a MGF origin because the abundance pattern synthesized in MGF ejecta remains uncertain theoretically and observationally. Nucleosynthesis in MGFs is sensitive to the initial thermodynamic conditions of the ejecta and its subsequent dynamical evolution, which are under active investigation \citep{Cehula+24, Bransgrove+2025}. More robust is the ability of MGFs to produce substantial $r$-process nuclei up to the second peak, with weak third peak production; the latter is a generic feature of decompressing neutron star crust across a wide range of initial conditions \citep{Goriely+2011, Patel+2025b}. Even if our current models disagree on the details, the enhancement observed in the UHCR abundances around the first and second $r$-process peak are thus consistent with a full or partial MGF origin. Nevertheless, definitive identification of the dominant source cannot be reached lacking higher fidelity modeling, further observations of MGF outflows, and improved statistical significance in UHCR measurements. Abundance measurements extending through the third $r$-process peak would also constrain the magnitude of any NSM contribution to the UHCR flux.

\section{Summary and Discussion}
\label{sec:conclude}

We evaluated the contributions of magnetar giant flares and neutron star mergers to the local ultra heavy cosmic ray flux by means of semi-analytic Galactic transport calculations. For a realistic distribution of sources in the Milky Way with presently available constraints on rates, energetics, and ejecta composition, we demonstrate that giant flares may contribute a persistent $r$-process element cosmic ray flux over Galactic history. The magnitude of this flux can be comparable to or exceed that from supernovae, depending on the giant flare rate and ion acceleration efficiency. The flux from neutron star mergers exhibits significant time-variability due to the low merger rate and severe propagation losses by nuclear spallation. Neutron star mergers therefore remain subdominant to supernovae through most of Galactic history, though a rare merger occurring within a few kiloparsecs of the Solar System may contribute to or even dominate the flux for a few million years.

We argue that the ultra heavy cosmic ray abundance data, in particular the breakdown of the volatility-based acceleration pattern coincident with an $r$-process element enhancement for $Z \gtrsim 40$, are consistent with a magnetar giant flare origin and in tension with a supernova origin. It is also possible that the anomaly be attributed to a combination of mergers and giant flares. This degeneracy may be broken by measurements with single-element resolution through the third $r$-process peak, which will be achieved by the upcoming TIGERISS mission \citep{Zober+2025}.

By identifying the source(s) of ultra heavy cosmic rays we obtain unique information into the history and properties of heavy nucleosynthetic events in our Galaxy over the last few million years. In particular, identification of $r$-process radioisotopes in the cosmic rays enables precision constraints on event ages. Comparison of abundances derived from meteorites with different cosmic ray exposure times also provides insight into the operation of persistent sources (giant flares) and time-variable sources (mergers) over an extended range in Galactic history \citep{Alexandrov+2022}. 

The spectra of ultra heavy cosmic rays encode signatures of their acceleration and transport processes, providing a distinct probe of the underlying microphysics. Transport properties are typically constrained with phenomenological models and global fits to high resolution data available for light nuclear species (e.g., \citealt{ Schroer+2021}). These species are predominantly accelerated by supernovae, introducing an intrinsic bias towards the properties of a single source class. Notably, the microphysical justification for transport parameters derived from phenomenological fits remains an open problem \citep{Kempski&Quataert2022}. 
Analogous exercises with ultra heavy species offer independent constraints to refine transport models; for instance these species should exhibit a gradual spectral break near $\varepsilon \sim 10^{12 \text{--} 13}~\rm eV~A^{-1}$ as their transport transitions from a spallation-limited regime reflecting the hard source spectra, to a softer diffusion-limited regime. The diagnostic potential of such features underscores the need for experimental access to resolved energy spectra and precise spallation cross sections for ultra heavy nuclei.

Our results are sensitive to the efficiency at which ultra heavy ions are accelerated at reverse shocks in the ejecta of magnetar giant flares and neutron star mergers. Future work should address the ejecta and shock dynamics of these outflows, as well as the microphysics of ultra heavy ion acceleration in collisionless shocks.

\begin{acknowledgments}
A.P. thanks Luca Comisso, Phillip Kempski, and Benedikt Schroer for helpful discussions.  This work was supported in part by the National Science Foundation (grant AST-2406637), NASA (grants 80NSSC22K0807, 80NSSC24K0408), and the Simons Foundation (grant 727700).  The Flatiron Institute is supported by the Simons Foundation.
\end{acknowledgments}

\appendix

\section{Stochastic Source Populations}
\label{sec:rsg}
The number of events in a source class over a time $t$ is estimated from a Poisson distribution with expected value $\mathcal{R}t$. The occurrence time of an event is sampled uniformly $t_0 \in [0, t]$.

Core-collapse SNe are concentrated in star-forming regions due to their short delay times after star formation. We use rejection sampling to generate a spatial distribution weighted by star formation density, which is described by a three dimensional spiral arm model calibrated using VLBI trigonometric parallaxes of molecular masers in the Milky Way \citep{Amend+2025, Reid+2019},
\be
\rho_{\rm sf}(\varrho, \varphi, z) \propto \exp(-z/z_{\rm  s})\exp(-\varrho/r_{\rm  s})\left( \frac{\varrho}{\varrho_{\rm  s}} \right)^2 \sum_{i=1}^5 \sigma_{{\rm arm},i}(\varrho, \varphi),
\label{eq:spiral}
\ee
where the surface density of the $i$'th arm $\sigma_{{\rm arm},i}$ is described in \citet[their Eq.~2]{Amend+2025}. We use a scale height $z_{\rm  s} = 0.15~\rm kpc$ and and length $\varrho_{\rm  s} = 4.0~\rm kpc$. Thermonuclear SNe experience longer delay times; their progenitors may therefore drift from their formation sites, resulting in a weaker spatial correlation with spiral arms (e.g., \citealt{Aramyan+2016}). We adopt a hybrid model for the spatial distribution of thermonuclear SNe, with one term proportional to the stellar mass and another proportional to star formation \citep[their Eq.~1]{Scannapieco&Bildsten2005}. Assuming a Galactic star formation rate $1~\msun~\rm yr^{-1}$ in recent history \citep{Elia+2022} and a stellar mass $M_\star =6\times 10^{10}M_\odot $ \citep{Licquia&Newman2015}, we estimate that these events should remain concentrated in star forming regions with probability $0.5$. The remaining events should be distributed according to the stellar mass density, for which we use an axisymmetric thick disk profile \citep{Miyamoto&Nagai1975},
\be
\rho_\star(\varrho, z) = \left( \frac{b_\star^2 M_\star}{4\pi} \right) \frac{a_\star  \varrho^2 + \left( 3\sqrt{z^2 + b_\star^2} + a_\star \right)\left( \sqrt{z^2 + b_\star^2} + a_\star \right)^2}{\left[ \varrho^2 + \left( \sqrt{z^2 + b_\star^2} + a \right)^2  \right]^{5/2} \left( z^2 + b_\star^2 \right)^{3/2}},
\label{eq:disk}
\ee
with scale parameters $a_\star = 2.0~\rm kpc$, and $b_\star =0.30~\rm kpc$ for the Milky Way.

Compact object binaries experience substantial natal kicks $\gtrsim 100~\rm km~s^{-1}$ during their evolution and typically merge after long delay times $\sim 1~\rm Gyr$, motivating a NSM population distributed approximately isotropically in the Galactic halo \citep{Wu+2019}. Evolving the current sample of Galactic neutron star binaries in the Galactic potential indeed predicts merger locations with radial offsets in agreement with short gamma ray burst data \citep{Gaspari+2024}. However, the study indicates that velocities of the systems are not isotropically oriented, in agreement with the dynamical orbital model of \citet{Amend+2025} which predicts smaller vertical offsets from the Galactic plane as compared to an isotropic distribution. We therefore employ a hybrid model where each merger location is sampled according to the stellar mass density (Eq.~\ref{eq:disk}) with probability $0.85$ and is otherwise isotropically oriented (i.e., with azimuthal and polar coordinates sampled uniformly) about the Galactic center with radius sampled from a lognormal probability distribution fit to the projected short gamma ray burst offsets compiled in \citet{Fong+2022}. The resulting cumulative distributions as a function of Galactocentric radius and vertical offset are similar to those predicted by the dynamical orbital model.

The short lifetimes of magnetars ($10^3 \text{--}10^4~\rm yr$) constrain their spatial distribution to track that of their progenitor core-collapse SNe. The location of MGFs is then well approximated by the spiral arm model (Eq.~\ref{eq:spiral}). As we demonstrate in Fig.~\ref{fig:sources}, this is in agreement with the locations of the current sample of Galactic magnetars \citep{Olausen&Kaspi2014}.\footnote{McGill magnetar catalog: \url{https://www.physics.mcgill.ca/~pulsar/magnetar/main.html}}

\section{Injection Spectrum}
\label{sec:injection}
Here we derive estimates for the power-law index $q$, maximum energy $E_{\rm max}$, and normalization $\mathcal{N}$ characterizing the nonthermal source spectrum $N_0(E)$ produced through diffusive shock acceleration (Eq.~\ref{eq:spectrum}).

Kinetic simulations suggest that, for strong shocks (fluid compression ratio $\chi = 4$), postshock magnetic fluctuations drift away from the shock front at the local Alfvén speed with respect to the background plasma \citep{Haggerty&Caprioli2020, Caprioli+2020}. This drift, or “postcursor”, enhances escape from the acceleration region, raising the effective fluid compression ratio felt by the particles $\Tilde{\chi} = \chi/(1 +\sqrt{2\chi\xi_{\rm B,2}})$ and steepening the spectrum from the canonical prediction of linear diffusive shock acceleration \citep{Diesing&Caprioli2021},
\be
q = \frac{\Tilde{\chi} + 2}{\Tilde{\chi}-1}.
\ee
Assuming a fraction $\xi_{\rm B,2} =10^{-2}$ of the ejecta kinetic energy is converted to downstream magnetic pressure, we find $q\approx 2.4$.

The maximum kinetic energy of a particle is estimated by comparing its diffusive confinement length around the shock front to the size of the accelerator,
\be
\frac{D_1}{v_{0}} \approx \alpha r_{\rm dec},
\label{eq:confine}
\ee
where $D_1 = c\lambda_{\rm L}/3$ is the Bohm diffusion coefficient, $r_{\rm dec} \approx (3M_{\rm ej}/4\pi \rho_0)^{1/3}$ is the characteristic deceleration radius of the ejecta, and $v_0$ is the upstream fluid velocity, and $\rho_0$ is the upstream density. The prefactor $\alpha \lesssim 0.1$ accounts for the fact that the true acceleration region behind the shock is typically less than $r_{\rm dec}$ and the cosmic ray pressure is not constant in the precursor \citep{Diesing2023}. The relativistic gyroradius is $\lambda_{\rm L} \simeq E/ZeB_1$, and the precursor magnetic field $B_1$ is determined assuming saturation of the non-resonant hybrid instability \citep{Bell2004},
\be
\frac{B_1^2}{8\pi} = \frac{3}{2}\xi_{\rm CR}\rho_0 v_{0}^3c^{-1}.
\ee
This gives a maximum energy,
\be
E_{\rm max} \approx 12\alpha Ze\xi_{\rm CR}^{1/2} \rho_0^{1/6} v_0^{5/2}c^{-3/2}M_{\rm ej}^{1/3}, 
\ee
which reduces to Eq.~\eqref{eq:Emax} for an assumed $\alpha = 0.05$. At the deceleration radius, the mean ejecta density $\rho_{\rm ej}$ becomes comparable to the ambient medium density $\rho_{\rm ext}$, such that $\rho_0 \approx \rho_{\rm ej} \approx \rho_{\rm ext}$. The estimate above thus serves as a characteristic limit for acceleration at either the reverse or forward shock.

A fraction $\xi_{\rm CR}$ of the total energy budget is transferred to a cosmic ray species such that the normalization $\mathcal{N}$ is set by,
\be
\xi_{\rm CR} = \frac{1}{E_{\rm ej}} \int^\infty_{0.1mc^2} E N_0(E) dE.
\ee

\section{Propagation Model}
\label{sec:decay}
The transport equation (Eq.~\ref{eq:transport}) with free escape boundary conditions $N(E, z=\pm H, t) = 0$ is solved with the Green's function,
\be
G(E, \bm{r}, t; E_0, \bm{r}_0, t_0) = \frac{e^{-\zeta}}{(4\pi \lambda)^{3/2}} \delta(E_0 - E - \epsilon)
 \exp\left[-\frac{(\varrho -\varrho_0)^2}{4\lambda}\right] \sum_{k = -\infty}^{\infty} (-1)^k \exp\left[-\frac{(z - z_k)^2}{4\lambda}\right],
 \label{eq:green}
\ee
where $z_k = (-1)^k z_0 + 2kH$ and the $k\neq 0$ terms are image source positions reflected about the vertical boundaries. For a propagation time $\tau = t - t_0 >0$ we can write the spallation interaction depth $\zeta = \int_0^\tau \Gamma_{\rm sp}dt$, propagation length $\lambda^{1/2}= (\int_0^\tau Ddt)^{1/2}$, and cumulative energy loss $\epsilon = \int_0^\tau\dot{E}dt$.

The total particle cooling rate entering Eq.~\ref{eq:transport} is the sum of the dominant energy loss channels,
\be
\dot{E} = \dot{E}_{\rm ion} + \dot{E}_{\rm Coulomb} + \dot{E}_\pi.
\ee
The following prescriptions for these channels are detailed in \citet{Mannheim&Schlickeiser1994}.
The cooling rate of relativistic ions as they ionize neutral gas in the ISM is given by,
\begin{equation}
    \dot{E}_{\text{ion}} = -1.82 \times 10^{-7} ~\mathrm{eV~s^{-1}}\, Z^2 \bar{n}_1 \left( \frac{2\beta^2}{2\beta^3 + \beta_0^3} \right)
    \begin{cases}
        1 + 1.85 \times 10^{-2} \ln\beta & \text{if } \beta \ge \beta_0 \\
        1 & \text{if } \beta < \beta_0
    \end{cases},
    \label{eq:ion}
\end{equation}
where $\beta_0 = 0.01$ and $\bar{n} \equiv \bar{n}_1~\rm cm^{-3}$ is the effective gas density sampled by particles as they diffuse over their propagation volume as described in Sec.~\ref{sec:model}. Energy is also lost through Coulomb upscattering of thermal electrons, 
\be
\dot{E}_{\rm Coulomb}
= - 3 \times 10^{-7}~\mathrm{eV~s^{-1}}\, Z^2 x_e\bar{n}_{1} \bigg(\frac{\beta^2}{6\times 10^{-9}T_4^{3/2} + \beta^3}\bigg),
\label{eq:coulomb}
\ee
where $x_e$ is the hydrogen ionization fraction in the ISM and $T_e \equiv 10^4T_4~\rm K$ is the temperature of the free electron gas. 

Nucleon--nucleon collisions produce pions; the effective energy transfer rate from a relativistic nucleon population to pions is,
\be
\dot{E}_{\rm \pi} = - 1.2\times10^{-17}~\mathrm{eV~s^{-1}}\,\gamma^{3/4} \bar{n}_1 \Theta(\gamma - 1.3) 
\begin{cases}
    (E/9.38 \times 10^8~\rm eV)^{7.64}, & \text{if }E < 7\times 10^8~\rm eV\\
    0.13(E/9.38 \times 10^8~\rm eV)^{0.53}, & \text{if }E > 7\times 10^8~\rm eV
\end{cases}.
\label{eq:pi}
\ee
Relativistic nuclei also fragment and pion produce at a rate $\Gamma_{\rm sp} = \bar{n} \sigma \beta c$. We use an empirical formula for the inelastic spallation cross section \citep{Letaw+1983},
\begin{eqnarray}
    \sigma = \sigma_0
    \begin{cases}
    1, & \text{if } \varepsilon > 2~\rm{GeV~A^{-1}} \\
    1-0.62\exp \left(\varepsilon  /200~\rm MeV~A^{-1}\right)\sin \left[ 10.9  \left(\varepsilon/\rm MeV~A^{-1}\right)^{-0.28} \right], & \text{if } \varepsilon < 2~\rm{GeV~A^{-1}}
    \end{cases},
\end{eqnarray}
where $\sigma_0 = 45A^{0.7}\left[1+1.6\times10^{-2}\sin(5.3 -2.6 \ln A)\right]~\rm mb$.

The solar modulation of particles with energy $E \lesssim 1~\rm GeV~A^{-1}$ can be described with a correction factor and energy shift to the local interstellar particle flux (Eq.~\ref{eq:N_obs}; \citealt{Gleeson&Axford1968}),
\be
f_\odot(E) = \frac{E(E+2mc^2)}{(E + \Phi)(E + \Phi + 2mc^2)},
\ee
with a mean energy $\Phi = 800Z~\rm MeV$ lost against the modulation potential, for which we have used $800~\rm MV$ to achieve a good fit for the proton and iron spectra (Fig.~\ref{fig:spectrum}).

\bibliography{mgf, mgf_CR}{}

@ARTICLE{Goriely+2011,
       author = {{Goriely}, S. and {Chamel}, N. and {Janka}, H.-T. and {Pearson}, J.~M.},
        title = "{The decompression of the outer neutron star crust and r-process nucleosynthesis}",
      journal = {\aap},
         year = 2011,
        month = jul,
       volume = {531},
          eid = {A78},
        pages = {A78},
          doi = {10.1051/0004-6361/201116897},
archivePrefix = {arXiv},
       eprint = {1105.2453},
 primaryClass = {astro-ph.SR},
       adsurl = {https://ui.adsabs.harvard.edu/abs/2011A&A...531A..78G}
}

@ARTICLE{Kaspi&Beloborodov2017,
       author = {{Kaspi}, Victoria M. and {Beloborodov}, Andrei M.},
        title = "{Magnetars}",
      journal = {\araa},
         year = 2017,
        month = aug,
       volume = {55},
       number = {1},
        pages = {261-301},
          doi = {10.1146/annurev-astro-081915-023329},
archivePrefix = {arXiv},
       eprint = {1703.00068},
 primaryClass = {astro-ph.HE},
       adsurl = {https://ui.adsabs.harvard.edu/abs/2017ARA&A..55..261K}
}

@ARTICLE{Prantzos+2020,
       author = {{Prantzos}, N. and {Abia}, C. and {Cristallo}, S. and {Limongi}, M. and {Chieffi}, A.},
        title = "{Chemical evolution with rotating massive star yields II. A new assessment of the solar s- and r-process components}",
      journal = {\mnras},
         year = 2020,
        month = jan,
       volume = {491},
       number = {2},
        pages = {1832-1850},
          doi = {10.1093/mnras/stz3154},
archivePrefix = {arXiv},
       eprint = {1911.02545},
 primaryClass = {astro-ph.GA},
       adsurl = {https://ui.adsabs.harvard.edu/abs/2020MNRAS.491.1832P}
}

@ARTICLE{Patel+2025b,
       author = {{Patel}, Anirudh and {Metzger}, Brian D. and {Goldberg}, Jared A. and {Cehula}, Jakub and {Thompson}, Todd A. and {Renzo}, Mathieu},
        title = "{r-process Nucleosynthesis and Radioactively Powered Transients from Magnetar Giant Flares}",
      journal = {\apj},
         year = 2025,
        month = jun,
       volume = {985},
       number = {2},
          eid = {234},
        pages = {234},
          doi = {10.3847/1538-4357/adceb7},
archivePrefix = {arXiv},
       eprint = {2501.17253},
 primaryClass = {astro-ph.HE},
       adsurl = {https://ui.adsabs.harvard.edu/abs/2025ApJ...985..234P}
}

@ARTICLE{Olausen&Kaspi2014,
       author = {{Olausen}, S.~A. and {Kaspi}, V.~M.},
        title = "{The McGill Magnetar Catalog}",
      journal = {\apjs},
         year = 2014,
        month = may,
       volume = {212},
       number = {1},
          eid = {6},
        pages = {6},
          doi = {10.1088/0067-0049/212/1/6},
archivePrefix = {arXiv},
       eprint = {1309.4167},
 primaryClass = {astro-ph.HE},
       adsurl = {https://ui.adsabs.harvard.edu/abs/2014ApJS..212....6O}
}

@ARTICLE{patel+25a,
       author = {{Patel}, Anirudh and {Metzger}, Brian D. and {Cehula}, Jakub and {Burns}, Eric and {Goldberg}, Jared A. and {Thompson}, Todd A.},
        title = "{Direct Evidence for r-process Nucleosynthesis in Delayed MeV Emission from the SGR 1806{\textendash}20 Magnetar Giant Flare}",
      journal = {\apjl},
         year = 2025,
        month = may,
       volume = {984},
       number = {1},
          eid = {L29},
        pages = {L29},
          doi = {10.3847/2041-8213/adc9b0},
archivePrefix = {arXiv},
       eprint = {2501.09181},
 primaryClass = {astro-ph.HE},
       adsurl = {https://ui.adsabs.harvard.edu/abs/2025ApJ...984L..29P}
}

@ARTICLE{ligo_cat_4_2025,
       author = {{The LIGO Scientific Collaboration} and {the Virgo Collaboration} and {the KAGRA Collaboration} and {Abac}, A.~G. and {Abouelfettouh}, I. and {Acernese}, F. and {Ackley}, K. and {Adamcewicz}, C. and {Adhicary}, S. and {Adhikari}, D. and {Adhikari}, N. and {Adhikari}, R.~X. and {Adkins}, V.~K. and {Afroz}, S. and {Agarwal}, D. and {Agathos}, M. and {Aghaei Abchouyeh}, M. and {Aguiar}, O.~D. and {Ahmadzadeh}, S. and {Aiello}, L. and {Ain}, A. and {Ajith}, P. and {Akutsu}, T. and {Albanesi}, S. and {Alfaidi}, R.~A. and {Al-Jodah}, A. and {All{\'e}n{\'e}}, C. and {Allocca}, A. and {Al-Shammari}, S. and {Altin}, P.~A. and {Alvarez-Lopez}, S. and {Amarasinghe}, O. and {Amato}, A. and {Amra}, C. and {Ananyeva}, A. and {Anderson}, S.~B. and {Anderson}, W.~G. and {Andia}, M. and {Ando}, M. and {Andrade}, T. and {Andr{\'e}s-Carcasona}, M. and {Andri{\'c}}, T. and {Anglin}, J. and {Ansoldi}, S. and {Antelis}, J.~M. and {Antier}, S. and {Aoumi}, M. and {Appavuravther}, E.~Z. and {Appert}, S. and {Apple}, S.~K. and {Arai}, K. and {Araya}, A. and {Araya}, M.~C. and {Arca Sedda}, M. and {Areeda}, J.~S. and {Argianas}, L. and {Aritomi}, N. and {Armato}, F. and {Armstrong}, S. and {Arnaud}, N. and {Arogeti}, M. and {Aronson}, S.~M. and {Arun}, K.~G. and {Ashton}, G. and {Aso}, Y. and {Assiduo}, M. and {Assis de Souza Melo}, S. and {Aston}, S.~M. and {Astone}, P. and {Attadio}, F. and {Aubin}, F. and {AultONeal}, K. and {Avallone}, G. and {Babak}, S. and {Badaracco}, F. and {Badger}, C. and {Bae}, S. and {Bagnasco}, S. and {Bagui}, E. and {Baiotti}, L. and {Bajpai}, R. and {Baka}, T. and {Baker}, T. and {Ball}, M. and {Ballardin}, G. and {Ballmer}, S.~W. and {Banagiri}, S. and {Banerjee}, B. and {Bankar}, D. and {Baptiste}, T.~M. and {Baral}, P. and {Barayoga}, J.~C. and {Barish}, B.~C. and {Barker}, D. and {Barman}, N. and {Barneo}, P. and {Barone}, F. and {Barr}, B. and {Barsotti}, L. and {Barsuglia}, M. and {Barta}, D. and {Bartoletti}, A.~M. and {Barton}, M.~A. and {Bartos}, I. and {Basak}, S. and {Basalaev}, A. and {Bassiri}, R. and {Basti}, A. and {Bates}, D.~E. and {Bawaj}, M. and {Baxi}, P. and {Bayley}, J.~C. and {Baylor}, A.~C. and {Baynard}, II, P.~A. and {Bazzan}, M. and {Bedakihale}, V.~M. and {Beirnaert}, F. and {Bejger}, M. and {Belardinelli}, D. and {Bell}, A.~S. and {Bellie}, D.~S. and {Bellizzi}, L. and {Beltran-Martinez}, D. and {Benoit}, W. and {Bentara}, I. and {Bentley}, J.~D. and {Ben Yaala}, M. and {Bera}, S. and {Bergamin}, F. and {Berger}, B.~K. and {Bernuzzi}, S. and {Beroiz}, M. and {Berry}, C.~P.~L. and {Bersanetti}, D. and {Bertolini}, A. and {Betzwieser}, J. and {Beveridge}, D. and {Bevilacqua}, G. and {Bevins}, N. and {Bhandare}, R. and {Bhatt}, R. and {Bhattacharjee}, D. and {Bhaumik}, S. and {Bhowmick}, S. and {Biancalana}, V. and {Bianchi}, A. and {Bilenko}, I.~A. and {Billingsley}, G. and {Binetti}, A. and {Bini}, S. and {Binu}, C. and {Birnholtz}, O. and {Biscoveanu}, S. and {Bisht}, A. and {Bitossi}, M. and {Bizouard}, M. -A. and {Blaber}, S. and {Blackburn}, J.~K. and {Blagg}, L.~A. and {Blair}, C.~D. and {Blair}, D.~G. and {Bobba}, F. and {Bode}, N. and {Boileau}, G. and {Boldrini}, M. and {Bolingbroke}, G.~N. and {Bolliand}, A. and {Bonavena}, L.~D. and {Bondarescu}, R. and {Bondu}, F. and {Bonilla}, E. and {Bonilla}, M.~S. and {Bonino}, A. and {Bonnand}, R. and {Booker}, P. and {Borchers}, A. and {Borhanian}, S. and {Boschi}, V. and {Bose}, S. and {Bossilkov}, V. and {Boudon}, A. and {Bozzi}, A. and {Bradaschia}, C. and {Brady}, P.~R. and {Branch}, A. and {Branchesi}, M. and {Braun}, I. and {Briant}, T. and {Brillet}, A. and {Brinkmann}, M. and {Brockill}, P. and {Brockmueller}, E. and {Brooks}, A.~F. and {Brown}, B.~C. and {Brown}, D.~D. and {Brozzetti}, M.~L. and {Brunett}, S. and {Bruno}, G. and {Bruntz}, R. and {Bryant}, J.},
        title = "{GWTC-4.0: Population Properties of Merging Compact Binaries}",
      journal = {arXiv e-prints},
         year = 2025,
        month = aug,
          eid = {arXiv:2508.18083},
        pages = {arXiv:2508.18083},
          doi = {10.48550/arXiv.2508.18083},
archivePrefix = {arXiv},
       eprint = {2508.18083},
 primaryClass = {astro-ph.HE},
       adsurl = {https://ui.adsabs.harvard.edu/abs/2025arXiv250818083T}
}

@ARTICLE{Mereghetti+05,
       author = {{Mereghetti}, S. and {G{\"o}tz}, D. and {von Kienlin}, A. and {Rau}, A. and {Lichti}, G. and {Weidenspointner}, G. and {Jean}, P.},
        title = "{The First Giant Flare from SGR 1806-20: Observations Using the Anticoincidence Shield of the Spectrometer on INTEGRAL}",
      journal = {\apjl},
         year = 2005,
        month = may,
       volume = {624},
       number = {2},
        pages = {L105-L108},
          doi = {10.1086/430669},
archivePrefix = {arXiv},
       eprint = {astro-ph/0502577},
 primaryClass = {astro-ph},
       adsurl = {https://ui.adsabs.harvard.edu/abs/2005ApJ...624L.105M}
}

@ARTICLE{Frederiks+07,
       author = {{Frederiks}, D.~D. and {Golenetskii}, S.~V. and {Palshin}, V.~D. and {Aptekar}, R.~L. and {Ilyinskii}, V.~N. and {Oleinik}, F.~P. and {Mazets}, E.~P. and {Cline}, T.~L.},
        title = "{Giant flare in SGR 1806-20 and its Compton reflection from the Moon}",
      journal = {Astronomy Letters},
         year = 2007,
        month = jan,
       volume = {33},
       number = {1},
        pages = {1-18},
          doi = {10.1134/S106377370701001X},
archivePrefix = {arXiv},
       eprint = {astro-ph/0612289},
 primaryClass = {astro-ph},
       adsurl = {https://ui.adsabs.harvard.edu/abs/2007AstL...33....1F}
}

@ARTICLE{Bransgrove+2025,
       author = {{Bransgrove}, Ashley and {Beloborodov}, Andrei M. and {Levin}, Yuri},
        title = "{Hyperactive Magnetar Eruptions: Giant Flares, Baryon Ejections, and FRBs}",
      journal = {arXiv e-prints},
         year = 2025,
        month = aug,
          eid = {arXiv:2508.13419},
        pages = {arXiv:2508.13419},
archivePrefix = {arXiv},
       eprint = {2508.13419},
 primaryClass = {astro-ph.HE},
       adsurl = {https://ui.adsabs.harvard.edu/abs/2025arXiv250813419B}
}

@misc {Lodders+2020,
      author = "Katharina Lodders",
      title = "Solar Elemental Abundances",
      year = "2020",
      month = "12",
      publisher = "Oxford University Press",
      doi = "10.1093/acrefore/9780190647926.013.145",
      url = "https://oxfordre.com/planetaryscience/view/10.1093/acrefore/9780190647926.001.0001/acrefore-9780190647926-e-145"
}

@ARTICLE{Boggs+2007,
       author = {{Boggs}, Steven E. and {Zoglauer}, A. and {Bellm}, E. and {Hurley}, K. and {Lin}, R.~P. and {Smith}, D.~M. and {Wigger}, C. and {Hajdas}, W.},
        title = "{The Giant Flare of 2004 December 27 from SGR 1806-20}",
      journal = {\apj},
         year = 2007,
        month = may,
       volume = {661},
       number = {1},
        pages = {458-467},
          doi = {10.1086/516732},
archivePrefix = {arXiv},
       eprint = {astro-ph/0611318},
 primaryClass = {astro-ph},
       adsurl = {https://ui.adsabs.harvard.edu/abs/2007ApJ...661..458B}
}

@ARTICLE{Hotokezaka+2018,
       author = {{Hotokezaka}, Kenta and {Beniamini}, Paz and {Piran}, Tsvi},
        title = "{Neutron star mergers as sites of r-process nucleosynthesis and short gamma-ray bursts}",
      journal = {International Journal of Modern Physics D},
         year = 2018,
        month = jan,
       volume = {27},
       number = {13},
          eid = {1842005},
        pages = {1842005},
          doi = {10.1142/S0218271818420051},
archivePrefix = {arXiv},
       eprint = {1801.01141},
 primaryClass = {astro-ph.HE},
       adsurl = {https://ui.adsabs.harvard.edu/abs/2018IJMPD..2742005H}
}

@ARTICLE{Wu+2019,
       author = {{Wu}, Meng-Ru and {Banerjee}, Projjwal and {Metzger}, Brian D. and {Mart{\'\i}nez-Pinedo}, Gabriel and {Aramaki}, Tsuguo and {Burns}, Eric and {Hailey}, Charles J. and {Barnes}, Jennifer and {Karagiorgi}, Georgia},
        title = "{Finding the Remnants of the Milky Way's Last Neutron Star Mergers}",
      journal = {\apj},
         year = 2019,
        month = jul,
       volume = {880},
       number = {1},
          eid = {23},
        pages = {23},
          doi = {10.3847/1538-4357/ab2593},
archivePrefix = {arXiv},
       eprint = {1905.03793},
 primaryClass = {astro-ph.HE},
       adsurl = {https://ui.adsabs.harvard.edu/abs/2019ApJ...880...23W}
}

@ARTICLE{Taylor+2005,
       author = {{Taylor}, G.~B. and {Gelfand}, J.~D. and {Gaensler}, B.~M. and {Granot}, J. and {Kouveliotou}, C. and {Fender}, R.~P. and {Ramirez-Ruiz}, E. and {Eichler}, D. and {Lyubarsky}, Y.~E. and {Garrett}, M. and {Wijers}, R.~A.~M.~J.},
        title = "{The Growth, Polarization, and Motion of the Radio Afterglow from the Giant Flare from SGR 1806-20}",
      journal = {\apjl},
         year = 2005,
        month = nov,
       volume = {634},
       number = {1},
        pages = {L93-L96},
          doi = {10.1086/491648},
archivePrefix = {arXiv},
       eprint = {astro-ph/0504363},
 primaryClass = {astro-ph},
       adsurl = {https://ui.adsabs.harvard.edu/abs/2005ApJ...634L..93T}
}

@ARTICLE{Granot+2006,
       author = {{Granot}, J. and {Ramirez-Ruiz}, E. and {Taylor}, G.~B. and {Eichler}, D. and {Lyubarsky}, Y.~E. and {Wijers}, R.~A.~M.~J. and {Gaensler}, B.~M. and {Gelfand}, J.~D. and {Kouveliotou}, C.},
        title = "{Diagnosing the Outflow from the SGR 1806-20 Giant Flare with Radio Observations}",
      journal = {\apj},
         year = 2006,
        month = feb,
       volume = {638},
       number = {1},
        pages = {391-396},
          doi = {10.1086/497680},
archivePrefix = {arXiv},
       eprint = {astro-ph/0503251},
 primaryClass = {astro-ph},
       adsurl = {https://ui.adsabs.harvard.edu/abs/2006ApJ...638..391G}
}

@ARTICLE{Gelfand+2005,
       author = {{Gelfand}, J.~D. and {Lyubarsky}, Y.~E. and {Eichler}, D. and {Gaensler}, B.~M. and {Taylor}, G.~B. and {Granot}, J. and {Newton-McGee}, K.~J. and {Ramirez-Ruiz}, E. and {Kouveliotou}, C. and {Wijers}, R.~A.~M.~J.},
        title = "{A Rebrightening of the Radio Nebula Associated with the 2004 December 27 Giant Flare from SGR 1806-20}",
      journal = {\apjl},
         year = 2005,
        month = nov,
       volume = {634},
       number = {1},
        pages = {L89-L92},
          doi = {10.1086/498643},
archivePrefix = {arXiv},
       eprint = {astro-ph/0503269},
 primaryClass = {astro-ph},
       adsurl = {https://ui.adsabs.harvard.edu/abs/2005ApJ...634L..89G}
}

@ARTICLE{Thompson&Duncan1995,
       author = {{Thompson}, Christopher and {Duncan}, Robert C.},
        title = "{The soft gamma repeaters as very strongly magnetized neutron stars - I. Radiative mechanism for outbursts}",
      journal = {\mnras},
         year = 1995,
        month = jul,
       volume = {275},
       number = {2},
        pages = {255-300},
          doi = {10.1093/mnras/275.2.255},
       adsurl = {https://ui.adsabs.harvard.edu/abs/1995MNRAS.275..255T}
}

@ARTICLE{Mazets+1979,
       author = {{Mazets}, E.~P. and {Golentskii}, S.~V. and {Ilinskii}, V.~N. and {Aptekar}, R.~L. and {Guryan}, Iu. A.},
        title = "{Observations of a flaring X-ray pulsar in Dorado}",
      journal = {\nat},
         year = 1979,
        month = dec,
       volume = {282},
       number = {5739},
        pages = {587-589},
          doi = {10.1038/282587a0},
       adsurl = {https://ui.adsabs.harvard.edu/abs/1979Natur.282..587M}
}

@ARTICLE{Hurley+1999,
       author = {{Hurley}, K. and {Cline}, T. and {Mazets}, E. and {Barthelmy}, S. and {Butterworth}, P. and {Marshall}, F. and {Palmer}, D. and {Aptekar}, R. and {Golenetskii}, S. and {Il'Inskii}, V. and {Frederiks}, D. and {McTiernan}, J. and {Gold}, R. and {Trombka}, J.},
        title = "{A giant periodic flare from the soft {\ensuremath{\gamma}}-ray repeater SGR1900+14}",
      journal = {\nat},
         year = 1999,
        month = jan,
       volume = {397},
       number = {6714},
        pages = {41-43},
          doi = {10.1038/16199},
archivePrefix = {arXiv},
       eprint = {astro-ph/9811443},
 primaryClass = {astro-ph},
       adsurl = {https://ui.adsabs.harvard.edu/abs/1999Natur.397...41H}
}

@ARTICLE{Palmer+2005,
       author = {{Palmer}, D.~M. and {Barthelmy}, S. and {Gehrels}, N. and {Kippen}, R.~M. and {Cayton}, T. and {Kouveliotou}, C. and {Eichler}, D. and {Wijers}, R.~A.~M.~J. and {Woods}, P.~M. and {Granot}, J. and {Lyubarsky}, Y.~E. and {Ramirez-Ruiz}, E. and {Barbier}, L. and {Chester}, M. and {Cummings}, J. and {Fenimore}, E.~E. and {Finger}, M.~H. and {Gaensler}, B.~M. and {Hullinger}, D. and {Krimm}, H. and {Markwardt}, C.~B. and {Nousek}, J.~A. and {Parsons}, A. and {Patel}, S. and {Sakamoto}, T. and {Sato}, G. and {Suzuki}, M. and {Tueller}, J.},
        title = "{A giant {\ensuremath{\gamma}}-ray flare from the magnetar SGR 1806 - 20}",
      journal = {\nat},
         year = 2005,
        month = apr,
       volume = {434},
       number = {7037},
        pages = {1107-1109},
          doi = {10.1038/nature03525},
archivePrefix = {arXiv},
       eprint = {astro-ph/0503030},
 primaryClass = {astro-ph},
       adsurl = {https://ui.adsabs.harvard.edu/abs/2005Natur.434.1107P}
}

@ARTICLE{Hurley+05,
       author = {{Hurley}, K. and {Boggs}, S.~E. and {Smith}, D.~M. and {Duncan}, R.~C. and {Lin}, R. and {Zoglauer}, A. and {Krucker}, S. and {Hurford}, G. and {Hudson}, H. and {Wigger}, C. and {Hajdas}, W. and {Thompson}, C. and {Mitrofanov}, I. and {Sanin}, A. and {Boynton}, W. and {Fellows}, C. and {von Kienlin}, A. and {Lichti}, G. and {Rau}, A. and {Cline}, T.},
        title = "{An exceptionally bright flare from SGR 1806-20 and the origins of short-duration {\ensuremath{\gamma}}-ray bursts}",
      journal = {\nat},
         year = 2005,
        month = apr,
       volume = {434},
       number = {7037},
        pages = {1098-1103},
          doi = {10.1038/nature03519},
archivePrefix = {arXiv},
       eprint = {astro-ph/0502329},
 primaryClass = {astro-ph},
       adsurl = {https://ui.adsabs.harvard.edu/abs/2005Natur.434.1098H}
}

@ARTICLE{Cehula+24,
       author = {{Cehula}, Jakub and {Thompson}, Todd A. and {Metzger}, Brian D.},
        title = "{Dynamics of baryon ejection in magnetar giant flares: implications for radio afterglows, r-process nucleosynthesis, and fast radio bursts}",
      journal = {\mnras},
         year = 2024,
        month = mar,
       volume = {528},
       number = {3},
        pages = {5323-5345},
          doi = {10.1093/mnras/stae358},
archivePrefix = {arXiv},
       eprint = {2311.05681},
 primaryClass = {astro-ph.HE},
       adsurl = {https://ui.adsabs.harvard.edu/abs/2024MNRAS.528.5323C}
}

@article{cowperthwaite2017electromagnetic,
  title={The electromagnetic counterpart of the binary neutron star merger LIGO/Virgo GW170817. II. UV, optical, and near-infrared light curves and comparison to kilonova models},
  author={Cowperthwaite, PS and Berger, Eric and Villar, VA and Metzger, BD and Nicholl, M and Chornock, R and Blanchard, PK and Fong, W f and Margutti, R and Soares-Santos, M and others},
  journal={The Astrophysical Journal Letters},
  volume={848},
  number={2},
  pages={L17},
  year={2017},
  publisher={IOP Publishing}
}

@ARTICLE{Gall+17,
       author = {{Gall}, Christa and {Hjorth}, Jens and {Rosswog}, Stephan and {Tanvir}, Nial R. and {Levan}, Andrew J.},
        title = "{Lanthanides or Dust in Kilonovae: Lessons Learned from GW170817}",
      journal = {\apjl},
         year = 2017,
        month = nov,
       volume = {849},
       number = {2},
          eid = {L19},
        pages = {L19},
          doi = {10.3847/2041-8213/aa93f9},
archivePrefix = {arXiv},
       eprint = {1710.05863},
 primaryClass = {astro-ph.HE},
       adsurl = {https://ui.adsabs.harvard.edu/abs/2017ApJ...849L..19G}
}

@ARTICLE{Blandford&Ostriker1978,
       author = {{Blandford}, R.~D. and {Ostriker}, J.~P.},
        title = "{Particle acceleration by astrophysical shocks.}",
      journal = {\apjl},
         year = 1978,
        month = apr,
       volume = {221},
        pages = {L29-L32},
          doi = {10.1086/182658},
       adsurl = {https://ui.adsabs.harvard.edu/abs/1978ApJ...221L..29B}
}

@ARTICLE{Bell1978a,
       author = {{Bell}, A.~R.},
        title = "{The acceleration of cosmic rays in shock fronts - I.}",
      journal = {\mnras},
         year = 1978,
        month = jan,
       volume = {182},
        pages = {147-156},
          doi = {10.1093/mnras/182.2.147},
       adsurl = {https://ui.adsabs.harvard.edu/abs/1978MNRAS.182..147B}
}

@INPROCEEDINGS{Axford+1977,
       author = {{Axford}, W.~I. and {Leer}, E. and {Skadron}, G.},
        title = "{The Acceleration of Cosmic Rays by Shock Waves}",
    booktitle = {International Cosmic Ray Conference},
         year = 1977,
       series = {International Cosmic Ray Conference},
       volume = {11},
        month = jan,
        pages = {132},
       adsurl = {https://ui.adsabs.harvard.edu/abs/1977ICRC...11..132A}
}

@ARTICLE{Krymskii1977,
       author = {{Krymskii}, G.~F.},
        title = "{A regular mechanism for the acceleration of charged particles on the front of a shock wave}",
      journal = {Akademiia Nauk SSSR Doklady},
         year = 1977,
        month = jun,
       volume = {234},
        pages = {1306-1308},
       adsurl = {https://ui.adsabs.harvard.edu/abs/1977DoSSR.234.1306K}
}

@ARTICLE{Kempski&Quataert2022,
       author = {{Kempski}, Philipp and {Quataert}, Eliot},
        title = "{Reconciling cosmic ray transport theory with phenomenological models motivated by Milky-Way data}",
      journal = {\mnras},
         year = 2022,
        month = jul,
       volume = {514},
       number = {1},
        pages = {657-674},
          doi = {10.1093/mnras/stac1240},
archivePrefix = {arXiv},
       eprint = {2109.10977},
 primaryClass = {astro-ph.HE},
       adsurl = {https://ui.adsabs.harvard.edu/abs/2022MNRAS.514..657K}
}

@ARTICLE{Jikei+2025,
       author = {{Jikei}, Taiki and {Groselj}, Daniel and {Sironi}, Lorenzo},
        title = "{Magnetic Field Amplification and Particle Acceleration in Weakly Magnetized Trans-relativistic Electron-ion Shocks}",
      journal = {arXiv e-prints},
         year = 2025,
        month = dec,
          eid = {arXiv:2512.03169},
        pages = {arXiv:2512.03169},
          doi = {10.48550/arXiv.2512.03169},
archivePrefix = {arXiv},
       eprint = {2512.03169},
 primaryClass = {astro-ph.HE},
       adsurl = {https://ui.adsabs.harvard.edu/abs/2025arXiv251203169J}
}

@article{Evoli+2020,
  title = {AMS-02 beryllium data and its implication for cosmic ray transport},
  author = {Evoli, Carmelo and Morlino, Giovanni and Blasi, Pasquale and Aloisio, Roberto},
  journal = {Phys. Rev. D},
  volume = {101},
  issue = {2},
  pages = {023013},
  numpages = {11},
  year = {2020},
  month = {Jan},
  publisher = {American Physical Society},
  doi = {10.1103/PhysRevD.101.023013},
  url = {https://link.aps.org/doi/10.1103/PhysRevD.101.023013}
}

@ARTICLE{Schroer+2021,
       author = {{Schroer}, Benedikt and {Evoli}, Carmelo and {Blasi}, Pasquale},
        title = "{Intermediate-mass and heavy Galactic cosmic-ray nuclei: The case of new AMS-02 measurements}",
      journal = {\prd},
         year = 2021,
        month = jun,
       volume = {103},
       number = {12},
          eid = {123010},
        pages = {123010},
          doi = {10.1103/PhysRevD.103.123010},
archivePrefix = {arXiv},
       eprint = {2102.12576},
 primaryClass = {astro-ph.HE},
       adsurl = {https://ui.adsabs.harvard.edu/abs/2021PhRvD.103l3010S}
}

@ARTICLE{Blasi2017,
       author = {{Blasi}, P.},
        title = "{On the spectrum of stable secondary nuclei in cosmic rays}",
      journal = {\mnras},
         year = 2017,
        month = oct,
       volume = {471},
       number = {2},
        pages = {1662-1670},
          doi = {10.1093/mnras/stx1696},
archivePrefix = {arXiv},
       eprint = {1707.00525},
 primaryClass = {astro-ph.HE},
       adsurl = {https://ui.adsabs.harvard.edu/abs/2017MNRAS.471.1662B}
}

@ARTICLE{Woosley+Heger2007,
       author = {{Woosley}, S.~E. and {Heger}, A.},
        title = "{Nucleosynthesis and remnants in massive stars of solar metallicity}",
      journal = {\physrep},
         year = 2007,
        month = apr,
       volume = {442},
       number = {1-6},
        pages = {269-283},
          doi = {10.1016/j.physrep.2007.02.009},
archivePrefix = {arXiv},
       eprint = {astro-ph/0702176},
 primaryClass = {astro-ph},
       adsurl = {https://ui.adsabs.harvard.edu/abs/2007PhR...442..269W}
}

@ARTICLE{Rauch+2009,
       author = {{Rauch}, B.~F. and {Link}, J.~T. and {Lodders}, K. and {Israel}, M.~H. and {Barbier}, L.~M. and {Binns}, W.~R. and {Christian}, E.~R. and {Cummings}, J.~R. and {de Nolfo}, G.~A. and {Geier}, S. and {Mewaldt}, R.~A. and {Mitchell}, J.~W. and {Schindler}, S.~M. and {Scott}, L.~M. and {Stone}, E.~C. and {Streitmatter}, R.~E. and {Waddington}, C.~J. and {Wiedenbeck}, M.~E.},
        title = "{Cosmic Ray origin in OB Associations and Preferential Acceleration of Refractory Elements: Evidence from Abundances of Elements $_{26}$Fe through $_{34}$Se}",
      journal = {\apj},
         year = 2009,
        month = jun,
       volume = {697},
       number = {2},
        pages = {2083-2088},
          doi = {10.1088/0004-637X/697/2/2083},
archivePrefix = {arXiv},
       eprint = {0906.2021},
 primaryClass = {astro-ph.HE},
       adsurl = {https://ui.adsabs.harvard.edu/abs/2009ApJ...697.2083R}
}

@ARTICLE{Epstein1980,
       author = {{Epstein}, R.~I.},
        title = "{The acceleration of interstellar grains and the composition of the cosmic rays}",
      journal = {\mnras},
         year = 1980,
        month = dec,
       volume = {193},
        pages = {723-729},
          doi = {10.1093/mnras/193.4.723},
       adsurl = {https://ui.adsabs.harvard.edu/abs/1980MNRAS.193..723E}
}

@ARTICLE{Garcia-Munoz+1979,
       author = {{Garcia-Munoz}, M. and {Simpson}, J.~A. and {Wefel}, J.~P.},
        title = "{The isotopes of neon in the galactic cosmic rays.}",
      journal = {\apjl},
         year = 1979,
        month = sep,
       volume = {232},
        pages = {L95-L99},
          doi = {10.1086/183043},
       adsurl = {https://ui.adsabs.harvard.edu/abs/1979ApJ...232L..95G}
}

@article{Cristofari+2025,
    author = "Cristofari, P. and Tatischeff, V. and Chabot, M.",
    title = "{Diffusive shock acceleration of dust grains at supernova remnants}",
    eprint = "2410.23190",
    archivePrefix = "arXiv",
    primaryClass = "astro-ph.HE",
    doi = "10.1051/0004-6361/202452436",
    journal = "Astron. Astrophys.",
    volume = "693",
    pages = "A145",
    year = "2025"
}

@article{Meyer+1997,
    author = "Meyer, Jean-Paul and Drury, Luke O'C. and Ellison, Donald C.",
    title = "{Galactic cosmic rays from supernova remnants: 1. A Cosmic ray composition controlled by volatility and mass to charge ratio}",
    eprint = "astro-ph/9704267",
    archivePrefix = "arXiv",
    doi = "10.1086/304599",
    journal = "Astrophys. J.",
    volume = "487",
    pages = "182--196",
    year = "1997"
}

@ARTICLE{Caprioli&Spitkovsky2014,
       author = {{Caprioli}, D. and {Spitkovsky}, A.},
        title = "{Simulations of Ion Acceleration at Non-relativistic Shocks. II. Magnetic Field Amplification}",
      journal = {\apj},
         year = 2014,
        month = oct,
       volume = {794},
       number = {1},
          eid = {46},
        pages = {46},
          doi = {10.1088/0004-637X/794/1/46},
archivePrefix = {arXiv},
       eprint = {1401.7679},
 primaryClass = {astro-ph.HE},
       adsurl = {https://ui.adsabs.harvard.edu/abs/2014ApJ...794...46C}
}

@ARTICLE{Morlino&Caprioli2012,
       author = {{Morlino}, G. and {Caprioli}, D.},
        title = "{Strong evidence for hadron acceleration in Tycho's supernova remnant}",
      journal = {\aap},
         year = 2012,
        month = feb,
       volume = {538},
          eid = {A81},
        pages = {A81},
          doi = {10.1051/0004-6361/201117855},
archivePrefix = {arXiv},
       eprint = {1105.6342},
 primaryClass = {astro-ph.HE},
       adsurl = {https://ui.adsabs.harvard.edu/abs/2012A&A...538A..81M}
}

@ARTICLE{Ackermann+2011,
       author = {{Ackermann}, M. and {Ajello}, M. and {Allafort}, A. and {Baldini}, L. and {Ballet}, J. and {Barbiellini}, G. and {Bastieri}, D. and {Belfiore}, A. and {Bellazzini}, R. and {Berenji}, B. and {Blandford}, R.~D. and {Bloom}, E.~D. and {Bonamente}, E. and {Borgland}, A.~W. and {Bottacini}, E. and {Brigida}, M. and {Bruel}, P. and {Buehler}, R. and {Buson}, S. and {Caliandro}, G.~A. and {Cameron}, R.~A. and {Caraveo}, P.~A. and {Casandjian}, J.~M. and {Cecchi}, C. and {Chekhtman}, A. and {Cheung}, C.~C. and {Chiang}, J. and {Ciprini}, S. and {Claus}, R. and {Cohen-Tanugi}, J. and {de Angelis}, A. and {de Palma}, F. and {Dermer}, C.~D. and {do Couto e Silva}, E. and {Drell}, P.~S. and {Dumora}, D. and {Favuzzi}, C. and {Fegan}, S.~J. and {Focke}, W.~B. and {Fortin}, P. and {Fukazawa}, Y. and {Fusco}, P. and {Gargano}, F. and {Germani}, S. and {Giglietto}, N. and {Giordano}, F. and {Giroletti}, M. and {Glanzman}, T. and {Godfrey}, G. and {Grenier}, I.~A. and {Guillemot}, L. and {Guiriec}, S. and {Hadasch}, D. and {Hanabata}, Y. and {Harding}, A.~K. and {Hayashida}, M. and {Hayashi}, K. and {Hays}, E. and {J{\'o}hannesson}, G. and {Johnson}, A.~S. and {Kamae}, T. and {Katagiri}, H. and {Kataoka}, J. and {Kerr}, M. and {Kn{\"o}dlseder}, J. and {Kuss}, M. and {Lande}, J. and {Latronico}, L. and {Lee}, S.-H. and {Longo}, F. and {Loparco}, F. and {Lott}, B. and {Lovellette}, M.~N. and {Lubrano}, P. and {Martin}, P. and {Mazziotta}, M.~N. and {McEnery}, J.~E. and {Mehault}, J. and {Michelson}, P.~F. and {Mitthumsiri}, W. and {Mizuno}, T. and {Monte}, C. and {Monzani}, M.~E. and {Morselli}, A. and {Moskalenko}, I.~V. and {Murgia}, S. and {Naumann-Godo}, M. and {Nolan}, P.~L. and {Norris}, J.~P. and {Nuss}, E. and {Ohsugi}, T. and {Okumura}, A. and {Orlando}, E. and {Ormes}, J.~F. and {Ozaki}, M. and {Paneque}, D. and {Parent}, D. and {Pesce-Rollins}, M. and {Pierbattista}, M. and {Piron}, F. and {Pohl}, M. and {Prokhorov}, D. and {Rain{\`o}}, S. and {Rando}, R. and {Razzano}, M. and {Reposeur}, T. and {Ritz}, S. and {Parkinson}, P.~M. Saz and {Sgr{\`o}}, C. and {Siskind}, E.~J. and {Smith}, P.~D. and {Spinelli}, P. and {Strong}, A.~W. and {Takahashi}, H. and {Tanaka}, T. and {Thayer}, J.~G. and {Thayer}, J.~B. and {Thompson}, D.~J. and {Tibaldo}, L. and {Torres}, D.~F. and {Tosti}, G. and {Tramacere}, A. and {Troja}, E. and {Uchiyama}, Y. and {Vandenbroucke}, J. and {Vasileiou}, V. and {Vianello}, G. and {Vitale}, V. and {Waite}, A.~P. and {Wang}, P. and {Winer}, B.~L. and {Wood}, K.~S. and {Yang}, Z. and {Zimmer}, S. and {Bontemps}, S.},
        title = "{A Cocoon of Freshly Accelerated Cosmic Rays Detected by Fermi in the Cygnus Superbubble}",
      journal = {Science},
         year = 2011,
        month = nov,
       volume = {334},
       number = {6059},
        pages = {1103},
          doi = {10.1126/science.1210311},
       adsurl = {https://ui.adsabs.harvard.edu/abs/2011Sci...334.1103A}
}

@ARTICLE{Aharonian+2019,
       author = {{Aharonian}, Felix and {Yang}, Ruizhi and {de O{\~n}a Wilhelmi}, Emma},
        title = "{Massive stars as major factories of Galactic cosmic rays}",
      journal = {Nature Astronomy},
         year = 2019,
        month = mar,
       volume = {3},
        pages = {561-567},
          doi = {10.1038/s41550-019-0724-0},
archivePrefix = {arXiv},
       eprint = {1804.02331},
 primaryClass = {astro-ph.HE},
       adsurl = {https://ui.adsabs.harvard.edu/abs/2019NatAs...3..561A}
}

@ARTICLE{Higdon&Lingenfelter2003,
       author = {{Higdon}, J.~C. and {Lingenfelter}, R.~E.},
        title = "{The Superbubble Origin of $^{22}$Ne in Cosmic Rays}",
      journal = {\apj},
         year = 2003,
        month = jun,
       volume = {590},
       number = {2},
        pages = {822-832},
          doi = {10.1086/375192},
       adsurl = {https://ui.adsabs.harvard.edu/abs/2003ApJ...590..822H}
}

@ARTICLE{Binns+2016,
       author = {{Binns}, W.~R. and {Israel}, M.~H. and {Christian}, E.~R. and {Cummings}, A.~C. and {de Nolfo}, G.~A. and {Lave}, K.~A. and {Leske}, R.~A. and {Mewaldt}, R.~A. and {Stone}, E.~C. and {von Rosenvinge}, T.~T. and {Wiedenbeck}, M.~E.},
        title = "{Observation of the $^{60}$Fe nucleosynthesis-clock isotope in galactic cosmic rays}",
      journal = {Science},
         year = 2016,
        month = may,
       volume = {352},
       number = {6286},
        pages = {677-680},
          doi = {10.1126/science.aad6004},
       adsurl = {https://ui.adsabs.harvard.edu/abs/2016Sci...352..677B}
}

@ARTICLE{Binns+2005,
       author = {{Binns}, W.~R. and {Wiedenbeck}, M.~E. and {Arnould}, M. and {Cummings}, A.~C. and {George}, J.~S. and {Goriely}, S. and {Israel}, M.~H. and {Leske}, R.~A. and {Mewaldt}, R.~A. and {Meynet}, G. and {Scott}, L.~M. and {Stone}, E.~C. and {von Rosenvinge}, T.~T.},
        title = "{Cosmic-Ray Neon, Wolf-Rayet Stars, and the Superbubble Origin of Galactic Cosmic Rays}",
      journal = {\apj},
         year = 2005,
        month = nov,
       volume = {634},
       number = {1},
        pages = {351-364},
          doi = {10.1086/496959},
archivePrefix = {arXiv},
       eprint = {astro-ph/0508398},
 primaryClass = {astro-ph},
       adsurl = {https://ui.adsabs.harvard.edu/abs/2005ApJ...634..351B}
}

@ARTICLE{Murphy+2016,
       author = {{Murphy}, R.~P. and {Sasaki}, M. and {Binns}, W.~R. and {Brandt}, T.~J. and {Hams}, T. and {Israel}, M.~H. and {Labrador}, A.~W. and {Link}, J.~T. and {Mewaldt}, R.~A. and {Mitchell}, J.~W. and {Rauch}, B.~F. and {Sakai}, K. and {Stone}, E.~C. and {Waddington}, C.~J. and {Walsh}, N.~E. and {Ward}, J.~E. and {Wiedenbeck}, M.~E.},
        title = "{Galactic Cosmic Ray Origins and OB Associations: Evidence from SuperTIGER Observations of Elements $_{26}$Fe through $_{40}$Zr}",
      journal = {\apj},
         year = 2016,
        month = nov,
       volume = {831},
       number = {2},
          eid = {148},
        pages = {148},
          doi = {10.3847/0004-637X/831/2/148},
archivePrefix = {arXiv},
       eprint = {1608.08183},
 primaryClass = {astro-ph.HE},
       adsurl = {https://ui.adsabs.harvard.edu/abs/2016ApJ...831..148M}
}

@ARTICLE{Binns+1989,
       author = {{Binns}, W.~R. and {Garrard}, T.~L. and {Gibner}, P.~S. and {Israel}, M.~H. and {Kertzman}, M.~P. and {Klarmann}, J. and {Newport}, B.~J. and {Stone}, E.~C. and {Waddington}, C.~J.},
        title = "{Abundances of Ultraheavy Elements in the Cosmic Radiation: Results from HEAO 3}",
      journal = {\apj},
         year = 1989,
        month = nov,
       volume = {346},
        pages = {997},
          doi = {10.1086/168082},
       adsurl = {https://ui.adsabs.harvard.edu/abs/1989ApJ...346..997B}
}

@article{Zober+2025,
    author = "Zober, Wolfgang V. and others",
    collaboration = "Tigeriss",
    title = "{Scientific Goals of the TIGERISS Mission}",
    doi = "10.22323/1.501.0164",
    journal = "PoS",
    volume = "ICRC2025",
    pages = "164",
    year = "2025"
}

@article{Walsh+2023,
    author = "Walsh, Nathan Elliot",
    title = "{SuperTIGER Abundances of Galactic Cosmic Rays for the Atomic Number (Z) Interval 40 to 56}",
    doi = "10.22323/1.444.0053",
    journal = "PoS",
    volume = "ICRC2023",
    pages = "053",
    year = "2023"
}

@ARTICLE{Reynolds2008,
       author = {{Reynolds}, S.~P.},
        title = "{Supernova remnants at high energy.}",
      journal = {\araa},
         year = 2008,
        month = sep,
       volume = {46},
        pages = {89-126},
          doi = {10.1146/annurev.astro.46.060407.145237},
       adsurl = {https://ui.adsabs.harvard.edu/abs/2008ARA&A..46...89R}
}

@ARTICLE{Donnelly+2012,
       author = {{Donnelly}, J. and {Thompson}, A. and {O'Sullivan}, D. and {Daly}, J. and {Drury}, L. and {Domingo}, V. and {Wenzel}, K.-P.},
        title = "{Actinide and Ultra-Heavy Abundances in the Local Galactic Cosmic Rays: An Analysis of the Results from the LDEF Ultra-Heavy Cosmic-Ray Experiment}",
      journal = {\apj},
         year = 2012,
        month = mar,
       volume = {747},
       number = {1},
          eid = {40},
        pages = {40},
          doi = {10.1088/0004-637X/747/1/40},
       adsurl = {https://ui.adsabs.harvard.edu/abs/2012ApJ...747...40D}
}

@ARTICLE{Alexandrov+2022,
       author = {{Alexandrov}, A.~B. and {Bagulya}, A.~V. and {Babaev}, P.~A. and {Chernyavsky}, M.~M. and {Gippius}, A.~A. and {Gorbunov}, S.~A. and {Grachev}, V.~M. and {Kalinina}, G.~V. and {Konovalova}, N.~S. and {Okateva}, N.~M. and {Polukhina}, N.~G. and {Rymzhanov}, R.~A. and {Starkov}, N.~I. and {Soe}, Than Naing and {Shchedrina}, T.~V. and {Volkov}, A.~E. and {Voronkov}, R.~A.},
        title = "{Insight into History of GCR Heavy Nuclei Fluxes by Their Tracks in Meteorites}",
      journal = {Physics of Atomic Nuclei},
         year = 2022,
        month = oct,
       volume = {85},
       number = {5},
        pages = {446-458},
          doi = {10.1134/S1063778822050039},
       adsurl = {https://ui.adsabs.harvard.edu/abs/2022PAN....85..446A}
}

@ARTICLE{Licquia&Newman2015,
       author = {{Licquia}, Timothy C. and {Newman}, Jeffrey A.},
        title = "{Improved Estimates of the Milky Way's Stellar Mass and Star Formation Rate from Hierarchical Bayesian Meta-Analysis}",
      journal = {\apj},
         year = 2015,
        month = jun,
       volume = {806},
       number = {1},
          eid = {96},
        pages = {96},
          doi = {10.1088/0004-637X/806/1/96},
archivePrefix = {arXiv},
       eprint = {1407.1078},
 primaryClass = {astro-ph.GA},
       adsurl = {https://ui.adsabs.harvard.edu/abs/2015ApJ...806...96L}
}

@ARTICLE{Elia+2022,
       author = {{Elia}, D. and {Molinari}, S. and {Schisano}, E. and {Soler}, J.~D. and {Merello}, M. and {Russeil}, D. and {Veneziani}, M. and {Zavagno}, A. and {Noriega-Crespo}, A. and {Olmi}, L. and {Benedettini}, M. and {Hennebelle}, P. and {Klessen}, R.~S. and {Leurini}, S. and {Paladini}, R. and {Pezzuto}, S. and {Traficante}, A. and {Eden}, D.~J. and {Martin}, P.~G. and {Sormani}, M. and {Coletta}, A. and {Colman}, T. and {Plume}, R. and {Maruccia}, Y. and {Mininni}, C. and {Liu}, S.~J.},
        title = "{The Star Formation Rate of the Milky Way as Seen by Herschel}",
      journal = {\apj},
         year = 2022,
        month = dec,
       volume = {941},
       number = {2},
          eid = {162},
        pages = {162},
          doi = {10.3847/1538-4357/aca27d},
archivePrefix = {arXiv},
       eprint = {2211.05573},
 primaryClass = {astro-ph.GA},
       adsurl = {https://ui.adsabs.harvard.edu/abs/2022ApJ...941..162E}
}

@ARTICLE{Scannapieco&Bildsten2005,
       author = {{Scannapieco}, Evan and {Bildsten}, Lars},
        title = "{The Type Ia Supernova Rate}",
      journal = {\apjl},
         year = 2005,
        month = aug,
       volume = {629},
       number = {2},
        pages = {L85-L88},
          doi = {10.1086/452632},
archivePrefix = {arXiv},
       eprint = {astro-ph/0507456},
 primaryClass = {astro-ph},
       adsurl = {https://ui.adsabs.harvard.edu/abs/2005ApJ...629L..85S}
}

@article{Adriani+2022,
  title = {Observation of Spectral Structures in the Flux of Cosmic-Ray Protons from 50 GeV to 60 TeV with the Calorimetric Electron Telescope on the International Space Station},
  author = {Adriani, O. and Akaike, Y. and Asano, K. and Asaoka, Y. and Berti, E. and Bigongiari, G. and Binns, W. R. and Bongi, M. and Brogi, P. and Bruno, A. and Buckley, J. H. and Cannady, N. and Castellini, G. and Checchia, C. and Cherry, M. L. and Collazuol, G. and Ebisawa, K. and Ficklin, A. W. and Fuke, H. and Gonzi, S. and Guzik, T. G. and Hams, T. and Hibino, K. and Ichimura, M. and Ioka, K. and Ishizaki, W. and Israel, M. H. and Kasahara, K. and Kataoka, J. and Kataoka, R. and Katayose, Y. and Kato, C. and Kawanaka, N. and Kawakubo, Y. and Kobayashi, K. and Kohri, K. and Krawczynski, H. S. and Krizmanic, J. F. and Maestro, P. and Marrocchesi, P. S. and Messineo, A. M. and Mitchell, J. W. and Miyake, S. and Moiseev, A. A. and Mori, M. and Mori, N. and Motz, H. M. and Munakata, K. and Nakahira, S. and Nishimura, J. and de Nolfo, G. A. and Okuno, S. and Ormes, J. F. and Ozawa, S. and Pacini, L. and Papini, P. and Rauch, B. F. and Ricciarini, S. B. and Sakai, K. and Sakamoto, T. and Sasaki, M. and Shimizu, Y. and Shiomi, A. and Spillantini, P. and Stolzi, F. and Sugita, S. and Sulaj, A. and Takita, M. and Tamura, T. and Terasawa, T. and Torii, S. and Tsunesada, Y. and Uchihori, Y. and Vannuccini, E. and Wefel, J. P. and Yamaoka, K. and Yanagita, S. and Yoshida, A. and Yoshida, K. and Zober, W. V.},
  collaboration = {CALET Collaboration},
  journal = {Phys. Rev. Lett.},
  volume = {129},
  issue = {10},
  pages = {101102},
  numpages = {8},
  year = {2022},
  month = {Sep},
  publisher = {American Physical Society},
  doi = {10.1103/PhysRevLett.129.101102},
  url = {https://link.aps.org/doi/10.1103/PhysRevLett.129.101102}
}

@article{Aguilar+2015,
  title = {Precision Measurement of the Proton Flux in Primary Cosmic Rays from Rigidity 1 GV to 1.8 TV with the Alpha Magnetic Spectrometer on the International Space Station},
  author = {Aguilar, M. and Aisa, D. and Alpat, B. and Alvino, A. and Ambrosi, G. and Andeen, K. and Arruda, L. and Attig, N. and Azzarello, P. and Bachlechner, A. and Barao, F. and Barrau, A. and Barrin, L. and Bartoloni, A. and Basara, L. and Battarbee, M. and Battiston, R. and Bazo, J. and Becker, U. and Behlmann, M. and Beischer, B. and Berdugo, J. and Bertucci, B. and Bigongiari, G. and Bindi, V. and Bizzaglia, S. and Bizzarri, M. and Boella, G. and de Boer, W. and Bollweg, K. and Bonnivard, V. and Borgia, B. and Borsini, S. and Boschini, M. J. and Bourquin, M. and Burger, J. and Cadoux, F. and Cai, X. D. and Capell, M. and Caroff, S. and Casaus, J. and Cascioli, V. and Castellini, G. and Cernuda, I. and Cerreta, D. and Cervelli, F. and Chae, M. J. and Chang, Y. H. and Chen, A. I. and Chen, H. and Cheng, G. M. and Chen, H. S. and Cheng, L. and Chou, H. Y. and Choumilov, E. and Choutko, V. and Chung, C. H. and Clark, C. and Clavero, R. and Coignet, G. and Consolandi, C. and Contin, A. and Corti, C. and Gil, E. Cortina and Coste, B. and Creus, W. and Crispoltoni, M. and Cui, Z. and Dai, Y. M. and Delgado, C. and Della Torre, S. and Demirk\"oz, M. B. and Derome, L. and Di Falco, S. and Di Masso, L. and Dimiccoli, F. and D\'{\i}az, C. and von Doetinchem, P. and Donnini, F. and Du, W. J. and Duranti, M. and D'Urso, D. and Eline, A. and Eppling, F. J. and Eronen, T. and Fan, Y. Y. and Farnesini, L. and Feng, J. and Fiandrini, E. and Fiasson, A. and Finch, E. and Fisher, P. and Galaktionov, Y. and Gallucci, G. and Garc\'{\i}a, B. and Garc\'{\i}a-L\'opez, R. and Gargiulo, C. and Gast, H. and Gebauer, I. and Gervasi, M. and Ghelfi, A. and Gillard, W. and Giovacchini, F. and Goglov, P. and Gong, J. and Goy, C. and Grabski, V. and Grandi, D. and Graziani, M. and Guandalini, C. and Guerri, I. and Guo, K. H. and Haas, D. and Habiby, M. and Haino, S. and Han, K. C. and He, Z. H. and Heil, M. and Hoffman, J. and Hsieh, T. H. and Huang, Z. C. and Huh, C. and Incagli, M. and Ionica, M. and Jang, W. Y. and Jinchi, H. and Kanishev, K. and Kim, G. N. and Kim, K. S. and Kirn, Th. and Kossakowski, R. and Kounina, O. and Kounine, A. and Koutsenko, V. and Krafczyk, M. S. and La Vacca, G. and Laudi, E. and Laurenti, G. and Lazzizzera, I. and Lebedev, A. and Lee, H. T. and Lee, S. C. and Leluc, C. and Levi, G. and Li, H. L. and Li, J. Q. and Li, Q. and Li, Q. and Li, T. X. and Li, W. and Li, Y. and Li, Z. H. and Li, Z. Y. and Lim, S. and Lin, C. H. and Lipari, P. and Lippert, T. and Liu, D. and Liu, H. and Lolli, M. and Lomtadze, T. and Lu, M. J. and Lu, S. Q. and Lu, Y. S. and Luebelsmeyer, K. and Luo, J. Z. and Lv, S. S. and Majka, R. and Ma\~n\'a, C. and Mar\'{\i}n, J. and Martin, T. and Mart\'{\i}nez, G. and Masi, N. and Maurin, D. and Menchaca-Rocha, A. and Meng, Q. and Mo, D. C. and Morescalchi, L. and Mott, P. and M\"uller, M. and Ni, J. Q. and Nikonov, N. and Nozzoli, F. and Nunes, P. and Obermeier, A. and Oliva, A. and Orcinha, M. and Palmonari, F. and Palomares, C. and Paniccia, M. and Papi, A. and Pauluzzi, M. and Pedreschi, E. and Pensotti, S. and Pereira, R. and Picot-Clemente, N. and Pilo, F. and Piluso, A. and Pizzolotto, C. and Plyaskin, V. and Pohl, M. and Poireau, V. and Postaci, E. and Putze, A. and Quadrani, L. and Qi, X. M. and Qin, X. and Qu, Z. Y. and R\"aih\"a, T. and Rancoita, P. G. and Rapin, D. and Ricol, J. S. and Rodr\'{\i}guez, I. and Rosier-Lees, S. and Rozhkov, A. and Rozza, D. and Sagdeev, R. and Sandweiss, J. and Saouter, P. and Sbarra, C. and Schael, S. and Schmidt, S. M. and von Dratzig, A. Schulz and Schwering, G. and Scolieri, G. and Seo, E. S. and Shan, B. S. and Shan, Y. H. and Shi, J. Y. and Shi, X. Y. and Shi, Y. M. and Siedenburg, T. and Son, D. and Spada, F. and Spinella, F. and Sun, W. and Sun, W. H. and Tacconi, M. and Tang, C. P. and Tang, X. W. and Tang, Z. C. and Tao, L. and Tescaro, D. and Ting, Samuel C. C. and Ting, S. M. and Tomassetti, N. and Torsti, J. and T\"urko\ifmmode \breve{g}\else \u{g}\fi{}lu, C. and Urban, T. and Vagelli, V. and Valente, E. and Vannini, C. and Valtonen, E. and Vaurynovich, S. and Vecchi, M. and Velasco, M. and Vialle, J. P. and Vitale, V. and Vitillo, S. and Wang, L. Q. and Wang, N. H. and Wang, Q. L. and Wang, R. S. and Wang, X. and Wang, Z. X. and Weng, Z. L. and Whitman, K. and Wienkenh\"over, J. and Wu, H. and Wu, X. and Xia, X. and Xie, M. and Xie, S. and Xiong, R. Q. and Xin, G. M. and Xu, N. S. and Xu, W. and Yan, Q. and Yang, J. and Yang, M. and Ye, Q. H. and Yi, H. and Yu, Y. J. and Yu, Z. Q. and Zeissler, S. and Zhang, J. H. and Zhang, M. T. and Zhang, X. B. and Zhang, Z. and Zheng, Z. M. and Zhuang, H. L. and Zhukov, V. and Zichichi, A. and Zimmermann, N. and Zuccon, P. and Zurbach, C.},
  collaboration = {AMS Collaboration},
  journal = {Phys. Rev. Lett.},
  volume = {114},
  issue = {17},
  pages = {171103},
  numpages = {9},
  year = {2015},
  month = {Apr},
  publisher = {American Physical Society},
  doi = {10.1103/PhysRevLett.114.171103},
  url = {https://link.aps.org/doi/10.1103/PhysRevLett.114.171103}
}

@ARTICLE{Fong+2022,
       author = {{Fong}, Wen-fai and {Nugent}, Anya E. and {Dong}, Yuxin and {Berger}, Edo and {Paterson}, Kerry and {Chornock}, Ryan and {Levan}, Andrew and {Blanchard}, Peter and {Alexander}, Kate D. and {Andrews}, Jennifer and {Cobb}, Bethany E. and {Cucchiara}, Antonino and {Fox}, Derek and {Fryer}, Chris L. and {Gordon}, Alexa C. and {Kilpatrick}, Charles D. and {Lunnan}, Ragnhild and {Margutti}, Raffaella and {Miller}, Adam and {Milne}, Peter and {Nicholl}, Matt and {Perley}, Daniel and {Rastinejad}, Jillian and {Escorial}, Alicia Rouco and {Schroeder}, Genevieve and {Smith}, Nathan and {Tanvir}, Nial and {Terreran}, Giacomo},
        title = "{Short GRB Host Galaxies. I. Photometric and Spectroscopic Catalogs, Host Associations, and Galactocentric Offsets}",
      journal = {\apj},
         year = 2022,
        month = nov,
       volume = {940},
       number = {1},
          eid = {56},
        pages = {56},
          doi = {10.3847/1538-4357/ac91d0},
archivePrefix = {arXiv},
       eprint = {2206.01763},
 primaryClass = {astro-ph.GA},
       adsurl = {https://ui.adsabs.harvard.edu/abs/2022ApJ...940...56F}
}
\bibliographystyle{aasjournal}

\end{document}